\documentclass[sigconf,9pt]{acmart}
\usepackage{algorithm}
\usepackage{algorithmic}
\usepackage{array}
\usepackage{balance}
\usepackage{enumitem}
\usepackage{colortbl}
\usepackage{soul}
\definecolor{LightCyan}{rgb}{0.88,1,1}
\definecolor{myred}{cmyk}{0, 0.7808, 0.4429, 0.2}
\definecolor{myteal}{cmyk}{0.1, 0.2, 0.043, 0.1}
\definecolor{myteal2}{cmyk}{0.72, 0, 0, 0.29}
\definecolor{white}{cmyk}{0, 0, 0, 0.6}

  {
    \begin{list}{$\bullet$}{%
      \setlength{\parsep}{0pt}%
      \setlength{\topsep}{0pt}%
      \setlength{\itemsep}{0pt}}}%
  {\end{list}
  }

\newcommand{\name}{{\em BuScope\ }}
\newcommand{\names}{{\em BuScope}}

\newcommand{\nameb}{{BuSCOPE~}} % for sections. otherwise caps are wrong

\begin{document}

\copyrightyear{2019} 
\acmYear{2019} 
\setcopyright{acmcopyright}
\acmConference[MobiSys '19]{The 17th Annual International Conference on Mobile Systems, Applications, and Services}{June 17--21, 2019}{Seoul, Republic of Korea}
\acmBooktitle{The 17th Annual International Conference on Mobile Systems, Applications, and Services (MobiSys '19), June 17--21, 2019, Seoul, Republic of Korea}
\acmPrice{15.00}
\acmDOI{10.1145/3307334.3326091}
\acmISBN{978-1-4503-6661-8/19/06}

\title[\nameb: Fusing Individual \& Aggregated Mobility Behavior]{\nameb: Fusing Individual \& Aggregated Mobility Behavior for ``Live'' Smart City Services}
\author[Meegahapola et al.]{Lakmal Meegahapola}
\affiliation{%
 \institution{Singapore Management University}
}
\email{lakmalm@smu.edu.sg}

\author[ ]{Thivya Kandappu}
\affiliation{%
 \institution{Singapore Management University}
}
\email{thivyak@smu.edu.sg}

\author[ ]{Kasthuri Jayarajah}
\affiliation{%
 \institution{Singapore Management University}
}
\email{kasthurij.2014@phdis.smu.edu.sg}

\author[ ]{Leman Akoglu}
\affiliation{%
 \institution{Carnegie Mellon University}
}
\email{lakoglu@andrew.cmu.edu}

\author[ ]{Shili Xiang}
\affiliation{%
 \institution{Institute for Infocomm Research}
}
\email{sxiang@i2r.a-star.edu.sg}

\author[ ]{Archan Misra}
\affiliation{%
 \institution{Singapore Management University}
}
\email{archanm@smu.edu.sg}

\begin{abstract}

While analysis of urban commuting data has a long and demonstrated history of providing useful insights into human mobility behavior, such analysis has been performed largely in offline fashion and to aid medium-to-long term urban planning. In this work, we demonstrate the power of applying predictive analytics on real-time mobility data, specifically the smart-card generated trip data of millions of public bus commuters in Singapore, to create two novel and ``live'' smart city services. The key analytical novelty in our work lies in combining two aspects of urban mobility: (a) conformity: which reflects the predictability in the aggregated flow of commuters along bus routes, and (b) regularity: which captures the repeated trip patterns of each individual commuter. We demonstrate that the fusion of these two measures of behavior can be performed at city-scale using our \name platform, and can be used to create two innovative smart city applications. The Last-Mile Demand Generator provides $O(mins)$ lookahead into the number of disembarking passengers at neighborhood bus stops; it achieves over 85\% accuracy in predicting such disembarkations by an ingenious combination of individual-level regularity with aggregate-level conformity. By moving driverless vehicles proactively to match this predicted demand, we can reduce wait times for disembarking passengers by over 75\%. Independently, the Neighborhood Event Detector uses outlier measures of currently operating buses to detect and spatiotemporally localize dynamic urban events, as much as 1.5 hours in advance, with a localization error of ~450 meters.

%To deal with the non-negligible volumes of real-time data generated (an average of XXX distinct bus boarding and disembarkation events daily), we describe various system optimization choices made to obtain the appropriate accuracy vs. computational overhead tradeoffs. Our results are very promising: (a) we show that we can predict the disembarkation volume, YYY minutes in advance, with an average error of less than XXX\% for individual bus stops, with the prediction error reducing to XXX\% when considering cumulative disembarkations  in XXX $m^2$ grids. Simulation studies show that this can translate into a XXX\% reduction in the mean waiting time experienced by last-mile commuters; and (b) by fusing outlier scores from multiple buses, we can not only achieve ZZZ\% recall, identifying XXX out of YYY representative events we consider, but localize such events to a mean distance error of 463.8 meters and a temporal margin of ~100 minutes.
\end{abstract}

% This section is to define CCS conepts and keywords


\maketitle
\sloppy

\section{Introduction} 
\label{sec:introduction}

Analysis of digitized urban transportation data, such as taxi location traces or bus commute records, has long been used for a variety of urban applications, such as building mobility models~\cite{Liu2009}, predicting likely future congestion hotspots~\cite{castro2012urban} or classifying land use~\cite{Pan2013}. In general, these applications operate in \emph{offline} fashion, analyzing historical data traces to generate policy-level outputs. In this paper, we instead focus on the opportunities of performing \emph{live} \& predictive analysis of such commuting data streams, to support soft-real time smart city operations.

We specifically focus on smart-card generated trip data for public buses in Singapore, where the vast majority of users tap-in and tap-out when boarding and disembarking from a bus, respectively, thereby providing $(origin, destination)$ records for individual trips.
Through careful analysis of a month's worth of anonymity-preserving smart-card generated bus trip data (a total of 108 million trips, taken by $\approx$ 5 million commuters), we show that \textbf{the vast majority of public bus trips are predictable, and driven by routine commuting patterns}. We shall show that such predictability manifests in two aspects: (a) \emph{individual-level regularity}, which allows us to predict an individual's point of disembarkation, as soon as she boards a bus, and (b) \emph{aggregate-level conformity}, which allows us to use historical commuting flows to identify a relatively small set of likely disembarkation points, even for commuters with no relevant prior travel history.

We emphasize the notion of a \emph{live} mobility analytics platform, which enables making operational decisions or generating neighborhood-level insights on streaming mobility data, with $O(mins)$ responsiveness. To support such soft-real time processing of the tens of thousands of passenger boarding and disembarkation events that occur city-wide per minute during peak commuting times, we shall develop \names, our server-based multi-threaded platform that continually generates updated per-passenger and per-bus insights. In particular, we shall use such insights for two novel applications:
\begin{itemize}[leftmargin=1em]

\item \emph{Last Mile Demand Generator (LM-Demand)}, which provides $O(mins)$ look-ahead into the number of passengers projected to disembark at different bus-stops. By using this demand projection to dynamically redirect the placement of unmanned Mobility-on-Demand (MoD) vehicles, we can tackle the important problem of improving the \emph{last-mile} commuting experience~\footnote{https://www.channelnewsasia.com/news/singapore/commentary-driverless-vehicles-reshape-singapore-smart-nation-9451258}.%\cite{XXX}.

\item \emph{Neighborhood Event Predictor (NE-Pred)}, which uses observed anomalous characteristics of `live' commuting flows to both predict and \emph{spatiotemporally localize} neighborhood-scale events. Identifying such events even before they start allow city authorities to intervene dynamically, such as dispatch traffic cops or adjust traffic light schedules.
\end{itemize}

Both of these exemplar applications are based on novel ways to harness this underlying predictability; while \emph{LM-Demand} combines individual-specific regularity and aggregate-level conformity to accurately predict disembarkation volumes at future downstream bus-stops, \emph{NE-Pred} uses bus-level outlier scores derived from the presence of \emph{irregular} commuters to derive the spatiotemporal coordinates of likely urban events. 

%Supporting ``live execution" of these analytics-driven services requires the tackling of several system-level challenges. First, the volume of generated transactional record (the tap-in and tap-out events) is reasonably charge--roughly XXX million distinct events, distributed across roughly XXX different bus service instances, daily. Moreover, our predictions need to be potentially updated and recomputed each time a bus crosses a bus stop, roughly XXX times in total per 16.5-hour (5.30am-midnight) day. Second, while all buses have cellular data connections, simply sending each transactional record instantly to a backend for analysis imposes non-negligible load, especially during peak hours. Third,  there are application-specific tradeoffs between spatiotemporal granularity of our analytics and the resulting prediction accuracy--e.g., the accuracy of disembarkation predictions are likely to be much higher when aggregated across a collection of neighborhood bus-stops. We shall demonstrate how to tackle these challenges through \names, our platform that combines edge-aggregation of records (on individual buses) with support for low-latency processing of such records at application-specific spatiotemporal granularity.

\textbf{Key Contributions:} Our key contributions in this paper are:
\begin{itemize}[leftmargin=1em]
\item \emph{Establishing the Predictability of Bus Commuting Patterns:} We first show that most trips have high predictability: this allows us to predict an individual's destination, given the originating bus stop, with high confidence, even when the specific trip has low support in past data. We shall subsequently introduce a hybrid model for disembarkation prediction combining both the individual-level regularity and aggregate flow-based conformity and show that this has high accuracy: we can predict the exact disembarkation with an accuracy of >85\% on both weekdays and weekends, with the mean location error in such prediction being ~480 meters (approx. 1-2 bus-stops) on weekdays. This hybrid technique is shown to achieve $\sim$ 30\% improvement in prediction accuracy over 2 alternative baseline methods that simply utilize aggregated historical data.

\item \emph{Demonstrating the Utility of Disembarkation Prediction for Last-Mile MoD Positioning:} Through our analysis, we show that we can additionally predict disembarkation bus-stop accurately an average of 9 bus-stops (2.89 kms) in advance. By feeding such predictions through a simulation model of neighborhood-level mobility, we show that predictive pre-placement of MoD vehicles (with capacities varying from $C=$1-3 passengers/vehicle) can reduce the waiting time experienced by disembarking commuters by over 75\%, to an average of less than 30 seconds, compared to ~2 mins for a reactive baseline ($C=1$).

\item \emph{Developing a Predictability-Driven Model for Event Detection \& Localization:} We develop a novel method for event/anomaly detection, which first computes a continually-updated outlier score for each operating bus based on the inherent predictability of its on-board commuters. The method then extrapolates this outlier score to downstream bus-stops, reflecting our hypothesis that events often attract commuters making non-regular trips. By then aggregating and spatiotemporally clustering such scores across (bus, bus-stop) combinations, we show that we can detect all 3 representative events with low average spatial error (463.8 meters), but also, on average, 100 minutes in advance of an event's start time. Moreover, we show that this approach of downstream extrapolation is superior to an alternative ``spot anomaly" technique that is based on changes in the disembarkation volume at individual bus stops: our approach typically identifies and localizes both macro and micro events 40-80 mins in advance of the `spot anomaly' method.

\item \emph{Operationalizing the Analytics through \names:} We present the design and implementation of \names, a platform that allows us to perform the predictive analytics outlined above in soft-real time, on underlying streaming data.  We show that \name is flexible enough to recompute the analytical insights, at both individual and bus-level specificity, very frequently for peak city-scale workloads---e.g., it incurs  17.33 msecs average latency to process each of $\approx$ 270,000 boarding and alighting transactions generated by  221,217 commuters on 3777 buses, during a typical weekday, 30 minute peak period.

\end{itemize}

While \emph{LM-Demand} and \emph{NE-Pred} are novel and innovative smart city applications, we believe that our broader contribution is in demonstrating the power of ``live" analytics on such underlying transportation transactional data, thereby potentially paving the way for public transport companies worldwide to make such pseudonymized data available in real-time.

\section{Dataset \& Applications} \label{sec:overview}

To provide a clearer understanding of the predictive analysis and the new applications that form the core of this paper, we first detail both our dataset as well as outline the high-level operation of \emph{LM-Demand} and \emph{NE-Pred}. 

\subsection{Dataset Description}
Our analysis and test of the developed analytics platform is based on the public transport smart-card data \footnote{https://www.ezlink.com.sg/} of 5.1 million commuters of Singapore. Because fares on the Singapore public transit system are distance-proportional and because smart-card fares are significantly lower than paying cash, the overwhelming majority of commuters utilize the smart-card to `tap-in' (while boarding) and `tap-out' (while disembarking), thus enabling the capture of the origin \& destination locations and timestamps of each journey. The dataset available to us consists of comprehensive records of the tap-in  and tap-out details of approximately 180 million trips made by commuters during the month of August, 2013. The data spans across 4913 bus stops and 153 MRT stations---for this paper, we focus solely on the 108 million bus journeys, i.e., those that start and end at bus-stops. The dataset is \emph{pseudonymized} and contains no explicit personally identifiable information (PII): each journey results in a unique commuter-specific entry with the fields described in Table~\ref{tab:dataset}, where the identifier $cid$ is unique for each smart-card. 
%It is important to note that the data allows us to both track the trip history of individual commuters (across all days and all buses/bus services) and the aggregated commuter dynamics on a single bus \emph{instance} (a specific trip made by a bus for a given service). %\am{In the table: (i) what tells us the `bus instance'; (ii) what is dir and regnum?}

%There are two salient aspects of Singapore's public transport system: (a) Fares are distance-proportional: accordingly, commuters have an incentive to `tap out' while disembarking, failing which they will be charged the maximum fare amount (by assuming that the user disembarks at the last stop); (b) While bus drives can issue single-use paper tickets for cash payments, the fares are much steeper (almost double) than those for smart-cards, with additional discounts for school children and senior citizens: accordingly, the overwhelming majority of commuters routinely utilize the smart-card. $<jid, cid, ptype, tmode, sernum, dir, regnum, bstop, astop, timestamp, dis, time>$.

\begin{table}[t]
\begin{small}

\vspace{-0.2 in}
\begin{center}
\resizebox{1\linewidth}{!}{
    \begin{tabular}{p{1.5cm} >{\arraybackslash}m{6cm}} 
       \hline
       \textbf{Attribute} & \textbf{Description} 
       \\ 
       \hline \hline 
       $jid$ & captures the unique ID of a journey (e.g., boarding at station X and alighting at station Y). \\
       %Note that, in Singapore, successful transfers made within 45 minutes since last alighting are considered as the same journey, hence, will be given the same ID. \\%This notion is captured in Fig.~\ref{fig:whatistrip}.\\
       $cid$ & represents the unique, randomly generated card ID of the commuter.\\
    %   $ptype$ & type of passenger categorized into 3 groups: (a) adult, (b) senior citizen and (c) student.\\
       $tmode$ & captures the mode of the transport -- (a) bus, (b) train and (c) light rail.\\
       $sernum$ & denotes the service number of the bus\\
       $dir$ & direction of the journey (to/from the origin hub)\\
       $regnum$ & bus instance ID\\
       $bstop, astop$ & ID of the boarding and alighting stops\\
       $timestamp$ & time of boarding \\
       $dis, time$ & total distance and sojourn time of the journey\\ \hline \hline 
    \end{tabular}}
\end{center}
\end{small}
\caption{Dataset Description.}
\vspace{-0.2in}
\label{tab:dataset}
\end{table}

% -- kas -- removing
% \begin{figure}
%   \includegraphics[width=\linewidth]{pictures/JourneyTrip.png}
%   \caption{Illustration of a Journey and its composition}
%   \label{fig:whatistrip}
% \end{figure}

\noindent \textbf{Common Definitions: } In anticipation of the analysis in Section~\ref{sec:travelpattern}, we define the following terms:

\noindent \textbf{Regularity/Support:} Similar to the traditional definition in data mining literature, the \emph{support} of a trip $jid$ by commuter $cid$ is defined as the fraction of total trips (by $cid$) with the same $bstop$ values (i.e., identical boarding stops), normalized by the total trips (in the entire dataset) involving $cid$. Note that the support definition may be time-interval specific, with a commuter's support defined separately, for example, for \emph{weekday peak period} vs. \emph{weekend off-peak period}.

\noindent \textbf{Confidence:} The \emph{confidence} of a trip $jid$ with a specific $(bstop,astop)$ tuple is defined as the probability of the user $cid$'s disembarking at bus-stop $astop$, \emph{given} that she has boarded at $bstop$--i.e., the ratio obtained by dividing the number of user $cid$'s trips with $(bstop,astop)$ as the source-destination pair, divided by the total number of user $cid$'s trips originating at $bstop$.

%%%%%%%%%%%%% FIGURE %%%%%%%%%%%%
% \begin{figure*}[!thb]
% \begin{center}
% % \begin{tabular}{cc}
% \includegraphics[width=6.8in]{pictures/LM-Demand-Graphic.jpg}
% % \end{tabular}
% \vspace{-0.1in}
% \caption{Disembarkation Prediction \& Last-Mile MoD Pre-placement}
% \vspace{-0.1in}
% \label{fig:scenario1}
% \end{center}
% \end{figure*}
% %%%%%%%%%%%%% FIG

\begin{figure*}[t]
\begin{center}
     \begin{minipage}[t]{0.6\textwidth}
         \includegraphics[width=4.0in,height=1.4in]{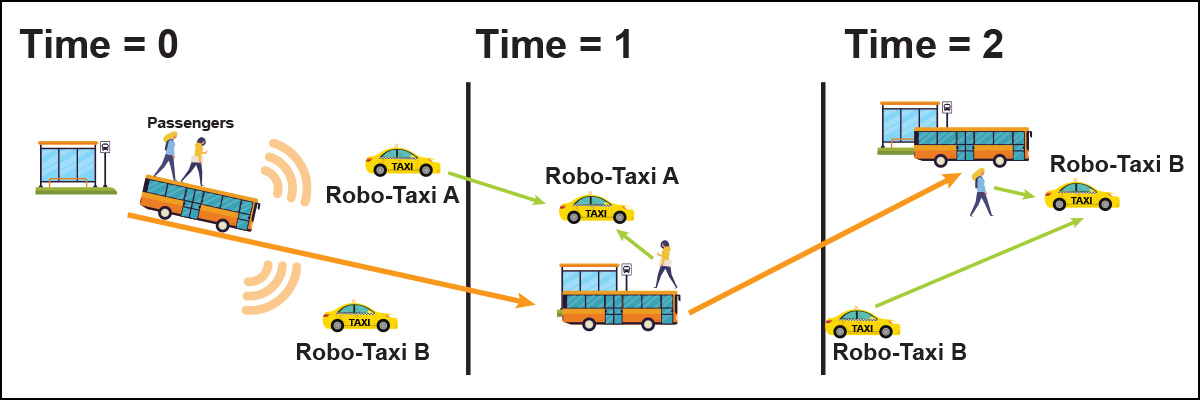}
         \caption{Disembarkation Prediction \& Last-Mile MoD Pre-placement}
         \label{fig:scenario1}
    \end{minipage}
    %\hfill
    \begin{minipage}[t]{0.3\textwidth}
        \includegraphics[width=2.5in, height=1.4in]{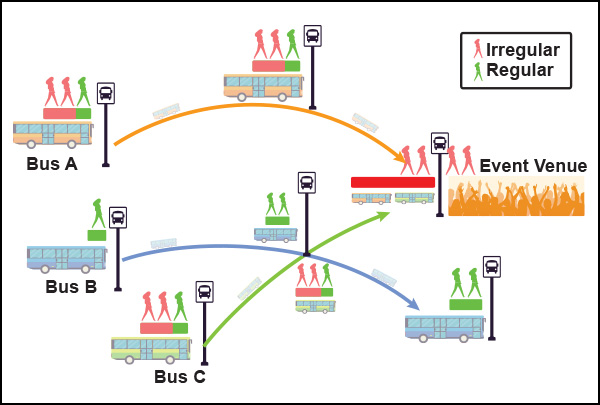}
        \caption{Mobility-Driven Event Detection \& Localization}
        \label{fig:scenario2}
    \end{minipage}
\end{center}
\vspace{-0.2 in}
\end{figure*}

\subsection{The Last-Mile Demand Prediction Application} While Singapore has an ambitious \emph{car-lite} vision that promotes extensive use of its excellent public transportation system, studies\footnote{https://www.todayonline.com/singapore/looking-ahead-2018-restoring-public-confidence-mrt-service-vital-steer-sporeans-away-cars} show that the overhead of the last-mile commute (the journey from/to the commuter's residence to/from the nearest bus stop) plays a big role in commuter reluctance to switch from a private car\cite{Krygsman2004}. Accordingly, Singapore is pursuing a vision of driverless MoD, where robo-taxis would ferry commuters to/from their doorstep to the nearest public transport node. 

A natural operational challenge in this setting is to maximize the utilization of such unmanned resources, and consequently \emph{minimize the waiting time of commuters}. Given a finite set of such MoD resources, the key to minimizing the waiting time (at least for the return commute) is to \emph{pre-position} the robo-taxis at the disembarkation points by anticipating the demand (the number of passengers disembarking at a bus stop at a future time instant). Figure~\ref{fig:scenario1} illustrates this concept, at a neighborhood level, with 3 different bus-stops and a 2 robo-taxis. Rather than allocate such MoD vehicles reactively (after passengers have disembarked and requested a ride) and cause passengers to wait, a smarter strategy would have proactively dispatched the robo-taxis to different bus-stops so that commuters find them ``magically" waiting as soon as they disembark--e.g., Robo-taxi A picks up the passenger at time $t=1$, while robo-taxi B moves to the third bus-stop to pick up the disembarking passenger at time $t=2$. A successful realization of this vision requires us to: (a) \emph{predict} the demand at each bus stop accurately, and (b) perform smart \emph{decision optimization} and proactively direct the robo-taxis to such predicted demand. In this work, we focus almost exclusively on the demand prediction aspect, and will show how \emph{LM-Demand}'s predictive analytics on such smart-card transactions can provide highly accurate estimates of the number of disembarking passengers, \emph{sufficiently in advance}. Of course, to illustrate the likely benefits of such prediction, we shall provide a comparative performance analysis of a straightforward MoD dispatch strategy, deferring the problem of algorithm design for predictive dispatch to future work.

\subsection{The Neighborhood Anomaly/Event Predictor} Large cities are highly dynamic, with potentially dozens of events (such as festivals, concerts and fairs) taking place in different city neighborhoods daily. City planners and urban agencies are very interested in detecting and tracking such events, to gain a better understanding of neighborhood dynamics, ascertain its \emph{livability} and also respond with timely interventions, such as dynamically adjusting transport network parameters (e.g, directionality of traffic lanes or duration of traffic lights) or deploying human resources (e.g., traffic officers) for better event management.

%%%%%%%%%%%%% FIGURE %%%%%%%%%%%%
% \begin{figure}[!thb]
% \begin{center}
% \includegraphics[width=0.5\textwidth]{pictures/NE-Pred-Graphic.jpg}
% \vspace{-0.1in}
% \caption{Mobility-Driven Event Detection \& Localization}
% \vspace{-0.1in}
% \label{fig:scenario2}
% \end{center}
% \end{figure}
% %%%%%%%%%%%%% FIG

A variety of approaches (e.g., using social media data~\cite{Sakaki:2010} or bike trip records~\cite{Zhang18}) have been proposed for such event detection. The challenge, of course, is to reliably isolate the contributory component of an event to such large-scale transactional data, from the daily dynamics of ``normal" mobility patterns. Our belief is that many such events cause residents to exhibit anomalous commuting patterns, and that some measure of anomaly aggregation, across the hundreds of geographically dispersed bus instances operational at any instance in a city, will provide a clear and reliable signal about the time and place of such underlying events. Bus usage data seems particularly appropriate for such prediction, as commuters heading to an event location are likely to board buses well in advance--e.g., 30 mins-1 hour before the start of an event. Figure~\ref{fig:scenario2} illustrates the high-level idea. We see an event in a city location, with `red' commuters denoting those exhibiting unusual travel patterns (e.g., travelling on routes that they don't normally use, or at hours not usually seen). Such `red' commuters are disproportionately present on buses heading \emph{towards} the event location, allowing the use of appropriate spatiotemporal clustering techniques to predictively localize the event. In this work, we shall focus on three aspects of this idea: (a) event detection: correctly declare the occurrence of such neighborhood-scale events; (b) event localization: accurately identify \emph{when} and \emph{where} such an event is happening; and (c) most innovatively, event prediction: forecast the start time of an event.

% \suhbsection{Our Approach at a Glance}
% Our eventual goal is to exploit the mutual reinforcement of regularity (user level) and conformity (aggregated flow level) to accurately predict (a) future user disembarkations and (b) current ridership levels. The workflow of this paper is given below:

% \begin{itemize}
    % \item \emph{Step 1:} Using the real-world, longitudinal fare-card data, we first uncover behavioral aspects of user commuting patterns. More specifically, (a) we discover aggregated commuter dynamics (i.e., collective observation of commuters board a bus from a transit hub alight at a shopping mall) and (b) individual travel specifics (such as longitudinal occurrences of boarding/alighting near home/work locations).
    % \item \emph{Step 2:} We then propose a hybrid approach that combines both (a) personalised and (b) flow based commuter dynamics to predict a commuter's disembarkation point given the boarding stop.
    % \item \emph{Step 3:} Next, we evaluate our approach, extensively -- we empirically analyse and present accuracy of (a) disembarkation prediction and (b) ridership estimation and (c) look-ahead distance (how many hops in advance we can predict).
    
% \end{itemize}

\section{Empirical Insights From City-scale Commuter Patterns} 
\label{sec:travelpattern}
Our exemplary applications and the overall design of \name are driven by a fundamental observation: \emph{the vast majority of bus trips undertaken by commuters, whether on weekdays or weekends, are in fact predictable}. Such predictability will enable us to predict (a) the number of disembarking passengers at downstream bus-stops (Section~\ref{sec:lmdemand}) or (b) the time \& location where an event will be held (Section~\ref{sec:anomaly}). In this section, we empirically demonstrate two key aspects of such predictability:
\begin{itemize}[leftmargin=1em]
    \item Predictability of a journey's destination--i.e., being able to infer where a passenger will disembark, given his embarkation context (such as the bus stop, bus service and time of boarding).
    \item Ridership mix in a bus--i.e., characterizing the mix of passengers on the bus who are exhibiting ``normal'' vs. ``abnormal" commuting patterns.
\end{itemize}
In addition, we also aim to understand the look-ahead time of such predictions.  We first look at the typical \emph{regularity} of commuting patterns, and uncover both individual-specific and aggregate-level properties that aid such predictions. 

%More specifically, we focus on the following questions:
%\begin{itemize}
%    \item What are the fundamental limits of `support' (i.e., frequency of historical boarding records of the same boarding point of interest) on predictability for his disembarkation? In particular, we focus on evaluating our ability to make predictions with high `confidence' even with low `support' values.
%    \item Can we utilize the global/aggregated commute trends to infer disembarkation points specially when we observe the user for the first time?
%\end{itemize}

\subsection{Person-Specific Commuting Regularity}
We first study the inherent \emph{regularity} of individual commuting patterns, viewing each trip as a $(origin, destination)$ tuple. Predicting the disembarkation location of a commuter is then driven by the conditional probability of alighting at bus-stop `Y', given the boarding bus-stop `X', and can be mined as an \emph{association rule} with a corresponding \emph{support} and \emph{confidence}. For example, assume that commuter $c$ has 100 trip records \emph{for a specific Context}, out of which 50 originate from bus stop `X', with 40 of those terminating at bus stop `Y'. Hence the rule $\{Boarding = X\} \Rightarrow \{Alighting = Y\}$ (interpreted as: if boarding stop=`X', then alighting stop=`Y') has a support of 50\% (=50/100) and confidence of  80\%(=40/50). More specifically, we define 4 different diurnal time windows\footnote{AM peak= (6-10:29am); AM off peak (10.30am-3:59pm), (c) PM peak (4pm-7.59pm), and (d) PM off peak (8pm-5:59am)}:  \{\emph{AM peak; AM off-peak; PM peak; PM off-peak}\}, corresponding to the four distinct service frequencies defined by the public bus services, each for two different day-of-week categories \{\emph{weekends}; \emph{weekdays}\}, resulting in 8 distinct contexts. Further, we consider two geographical areas in Singapore: one in the \emph{Central Business District} (CBD), and one in the more residential, \emph{Non- Central Business District} (NCBD) to capture varying dynamics of bus usage behavior.

%%%
\begin{figure}[!thb]
\begin{center}
\includegraphics[width=0.4\textwidth]{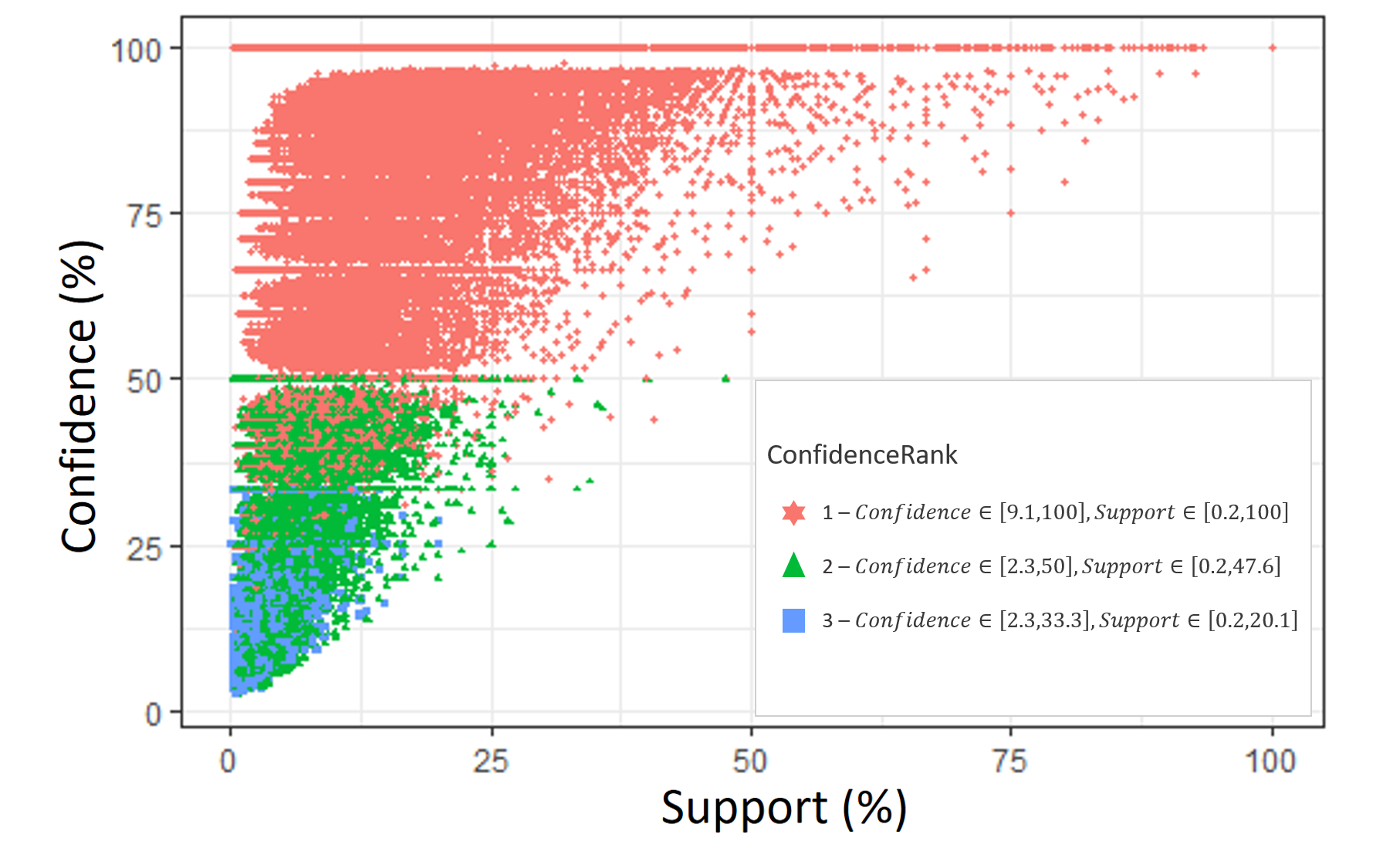}
\caption{Spread of Support and Confidence for various Confidence Ranks}
\label{fig:supconspread}
\vspace{-0.1in}
\end{center}
\vspace{-0.15in}
\end{figure}

In Figure~\ref{fig:supconspread}, we provide a scatterplot of the top-3 \emph{confidence} values, along with the corresponding \emph{support} for each trip (O-D pair) observed across all users. We employ a \emph{Common Path Re-identification} technique, whereby \emph{support} is defined based on (origin, destination) end-points and not just bus routes, as one O-D pair of bus-stops may often lie on the common route of multiple bus services. By aggregating trips made on such different services into a common (O,D) pair, we can improve both the \emph{support} and \emph{confidence} of individual journeys. %As is common with a well-connected public transport network (such as the one considered in this work), the network consists of multiple services that overlap with others in that they serve multiple common bus stops, along their route. As our $O,D$ pairs are service-dependent, we exploit this spatial structure to further optimize the confidence for the commuters with low support. We first identify set of bus services that traverse through a common path. The user's confidence for a particular alighting/boarding (boarding at `X, alighting at `Y' using service `b') in a given context is then recomputed by adding the all the confidence values corresponding to other bus services that traverse through `X' and `Y'.

\begin{figure*}[t]
\begin{center}
     \begin{minipage}[t]{0.33\textwidth}
         \includegraphics[width=2.3in,height=1.3in]{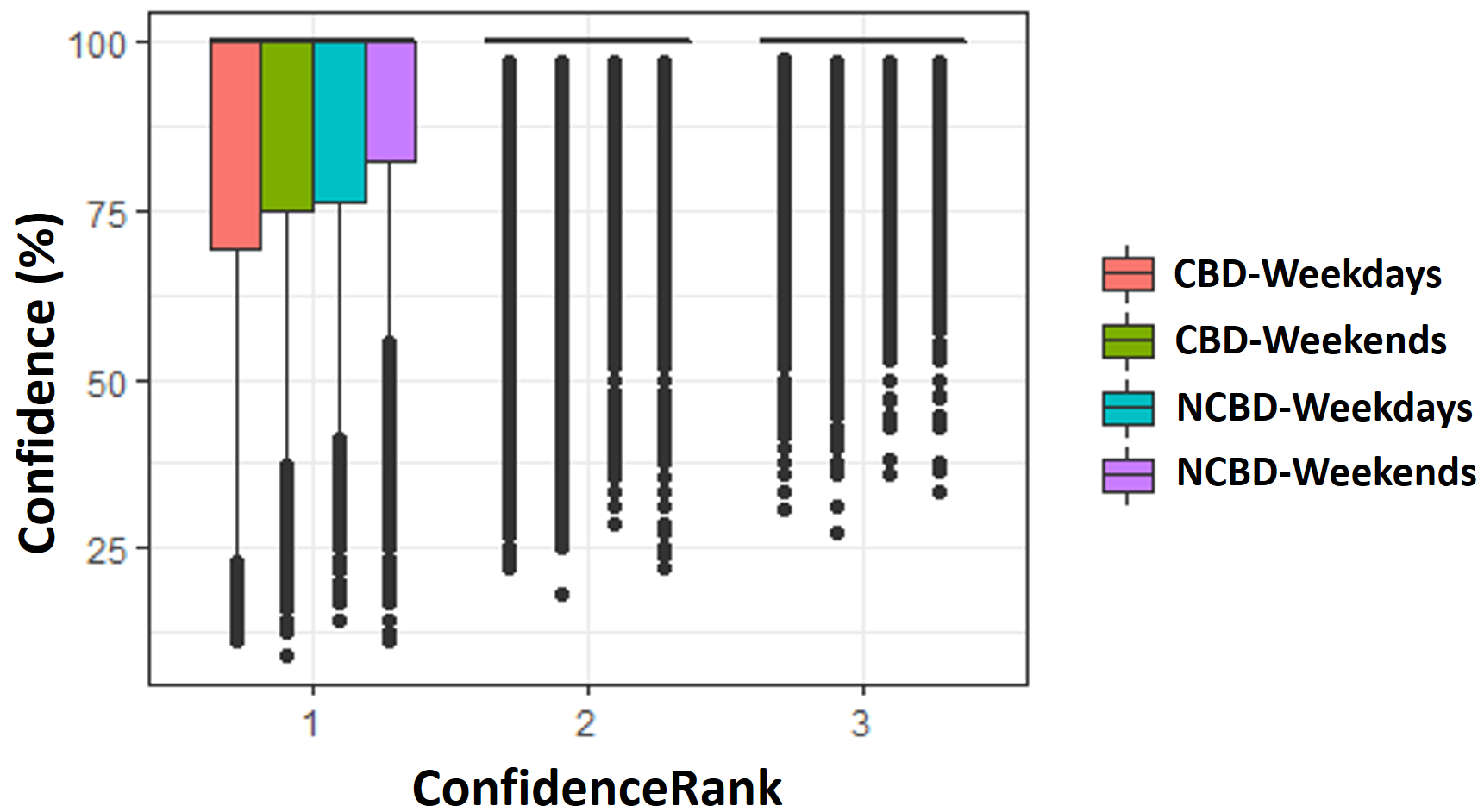}
         \caption{Spatiotemporal variability of confidence of personalized predictions}
         \label{fig:boxplot}
    \end{minipage}
    \hfill
    \begin{minipage}[t]{0.33\textwidth}
        \includegraphics[width=2.3in,height=1.3in]{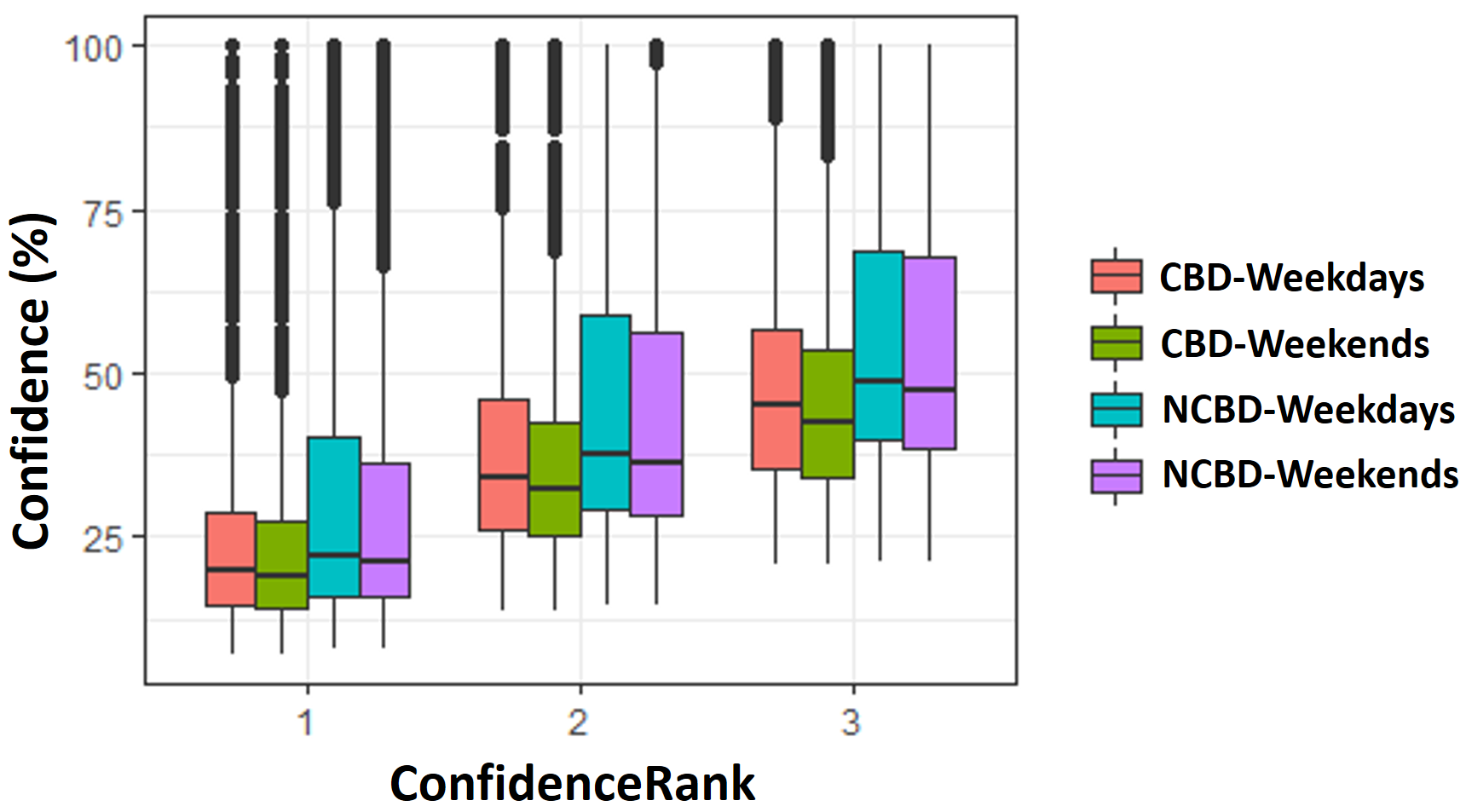}
        \caption{Spatiotemporal variability of confidence of flow-based predictions}
        \label{fig:boxplot2}
    \end{minipage}
    \begin{minipage}[t]{0.33\textwidth}
        \includegraphics[width=2.4in,height=1.3in]{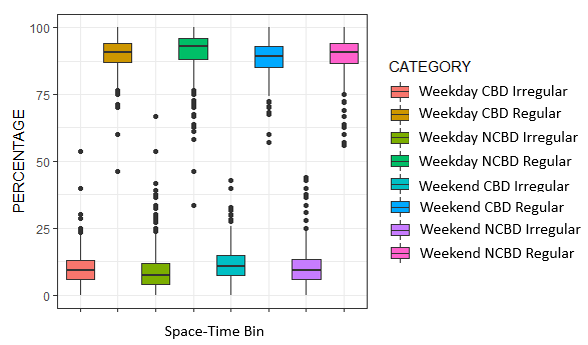}
        \caption{Fraction of \emph{Irregular} Passengers}
        \label{fig:boxplotfip}
    \end{minipage}
\end{center}
\vspace{-0.15in}
\end{figure*}

% \am{We should add a couple of lines of text how common paths on different buses are used to increase support.} 
We observe that the \emph{confidence} increases as support grows progressively larger--i.e., if a commuter has a past history of undertaking a particular journey repeatedly (e.g., home$\rightarrow$work), then we are more certain that he/she will alight at her workplace when starting a future journey from home. However, even for low support values (support<5\%), the confidence values are quite high (often exceeding 85\%). This result would suggest that \textbf{predicting disembarkation using such individualized travel history would suffice for users for whom there is even a modest past history of similar trips}. In Figure~\ref{fig:boxplot}, we plot the CDF of confidence for the top-3 (x-axis) most likely alighting destinations for different (O,D) pairs and space (CBD vs. NCBD) and time (weekdays vs. weekends) bins. We see that individual-level behavior is highly deterministic-- in the vast majority of cases, top-3 disembarkation predictions achieve $\approx$100\% confidence indicating that, for most originating destinations, a commuter disembarks at one of most 2-3 bus stops.

% \begin{figure}[ht]
%   \includegraphics[width =0.4\textwidth]{pictures/139FlowVisualizationV3.png}
%   \caption{Illustration of Transition (Disembarkation) Probabilities along a typical Bus Service ``139'' in Direction 1}
%   \label{fig:chorddiagramroute130}
% \end{figure}

\subsection{Generic Commuting Trend and Flow Based Patterns}
To additionally capture the fraction of bus passengers who do not have enough ``support" from past trips to make an individualized prediction, we next examine the overall aggregated \emph{flow-level} behavior of commuters. A significant amount of past literature on urban mobility has utilized such flow-level statistics. Our hypothesis is that some degree of prediction about an individual's likely disembarkation bus-stop may be gleaned by observing aggregate flow-level transition probabilities--i.e., by asking, what fraction, of the total number of individuals boarding at bus-stop `X', are observed to disembark at bus-stop `Y'? In Figure~\ref{fig:boxplot2}, we plot the CDF of confidence (per space-time bin) for the top-3 highly probable disembarkation at bus service level.%, for different space-time bins. 

%% kas -- moved up-- don't need this
% \begin{figure*}[!ht]
% \begin{center}
%      \begin{minipage}[t]{0.45\textwidth}
%          \includegraphics[width=3.5in,height=2in]{pictures/AggregateConfidenceRankPlotFLOW.png}
%          \caption{Variability in confidence (across top-3 disembarkation points) for Flow Based Predictions.}  \label{fig:boxplot2}
%     \end{minipage}
%     \hfill
%     \begin{minipage}[t]{0.45\textwidth}
%          {\includegraphics[width=3.5in]{pictures/139FlowVisualizationV2.png}
%          \caption{Illustrating Transition (Disembarkation) Probabilities across Typical Bus Service (``130'')}
%          \label{fig:chorddiagramroute130}
%         }
%     \end{minipage}
% \end{center}
% \end{figure*}

In general, we observe that, on average, given a source bus stop `X', the \emph{confidence} that embarking passengers will disembark at one of the top-3 probable locations is close to 50\%, even though the average number of stops along any service route is relatively larger, i.e., $\approx$ 49.77. Through further fine-grained analysis (details omitted due to space limits), we found that, for the vast majority of bus routes, these 3 stops are often common across different values of `X'. In other words, most bus routes had a few (usually 3-4) \emph{sink nodes}, which see a high volume of disembarking passengers, even though the passengers board the bus at a variety of bus-stops. As an illustration of this, Figure~\ref{fig:chorddiagramroute130} uses a ``chord diagram'' (see~\cite{colpaert2016public}) of both directions of the bus route ``139''. We can see the existence of a clear set of \emph{sink} nodes (e.g., bus stops 13099-- major residential estate and 7517 -- shopping district) that witness a disproportionately large volume of disembarking flows. 

%In additional analyses (not germane to this paper), we corroborated that these sink nodes match our intuitive expectation and almost always map to locations with certain key services, such as key transfer stations on the commuter rail network, major shopping malls or public arenas. Our empirical insights are important as they suggest that it \emph{may} be useful to probabilistically predict that irregular commuters (with inadequate past history) will 

\subsection{Typical Per-Bus Passenger Profiles }
Figure~\ref{fig:boxplotfip} plots the fraction of commuters (daily, over the observation period of 30 days) whose trip can be characterized as ``irregular''--i.e., for which, there are no records of similar past trips (for this plot, we use the first 3 weeks as training data, and the last week as the test period.) We see that, across the entire city, the fraction of such irregular trips is low, but not negligible (about 15\%), implying that ignoring the impact of such irregular commuters may lead to misleading predictions. Moreover, the fraction of such irregular users on any bus is observed to be high only rarely (in less than 5\% of the bus instances captured in our 1-month data), suggesting that \textbf{this may be a feature potentially indicative of  unusual events occurring along or near the corresponding bus route}.

\section{The \nameb System} 
\label{sec:system}

%\begin{figure*}[tbh]
%\begin{center}
%\includegraphics[width=4in, height=2.5in]{pictures/BuSCOPE-System.jpg}
%\caption{Functional Components of \name}
%\label{fig:systemarch}
%\vspace{-0.3in}
%\end{center}
%\end{figure*}

\begin{figure*}[t]
\begin{center}
     \begin{minipage}[t]{0.4\textwidth}
         \includegraphics[width=2.1in,height=2.0in]{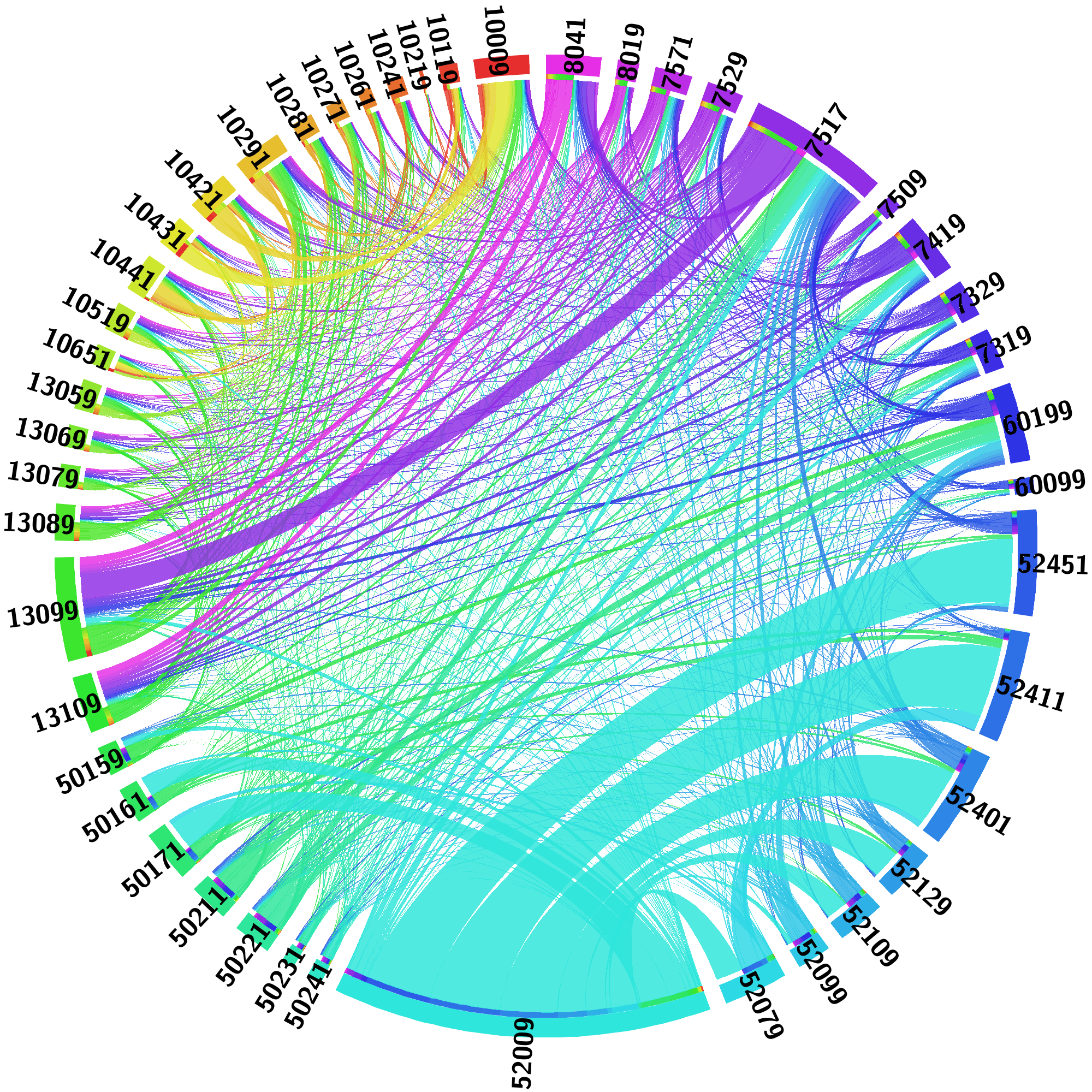}
         \caption{Visualizing Disembarkation Probabilities for Bus Service ``139'' (Direction 1)}
         \label{fig:chorddiagramroute130}
    \end{minipage}
    %\hfill
    \begin{minipage}[t]{0.55\textwidth}
        \includegraphics[width=4in,height=2.2in]{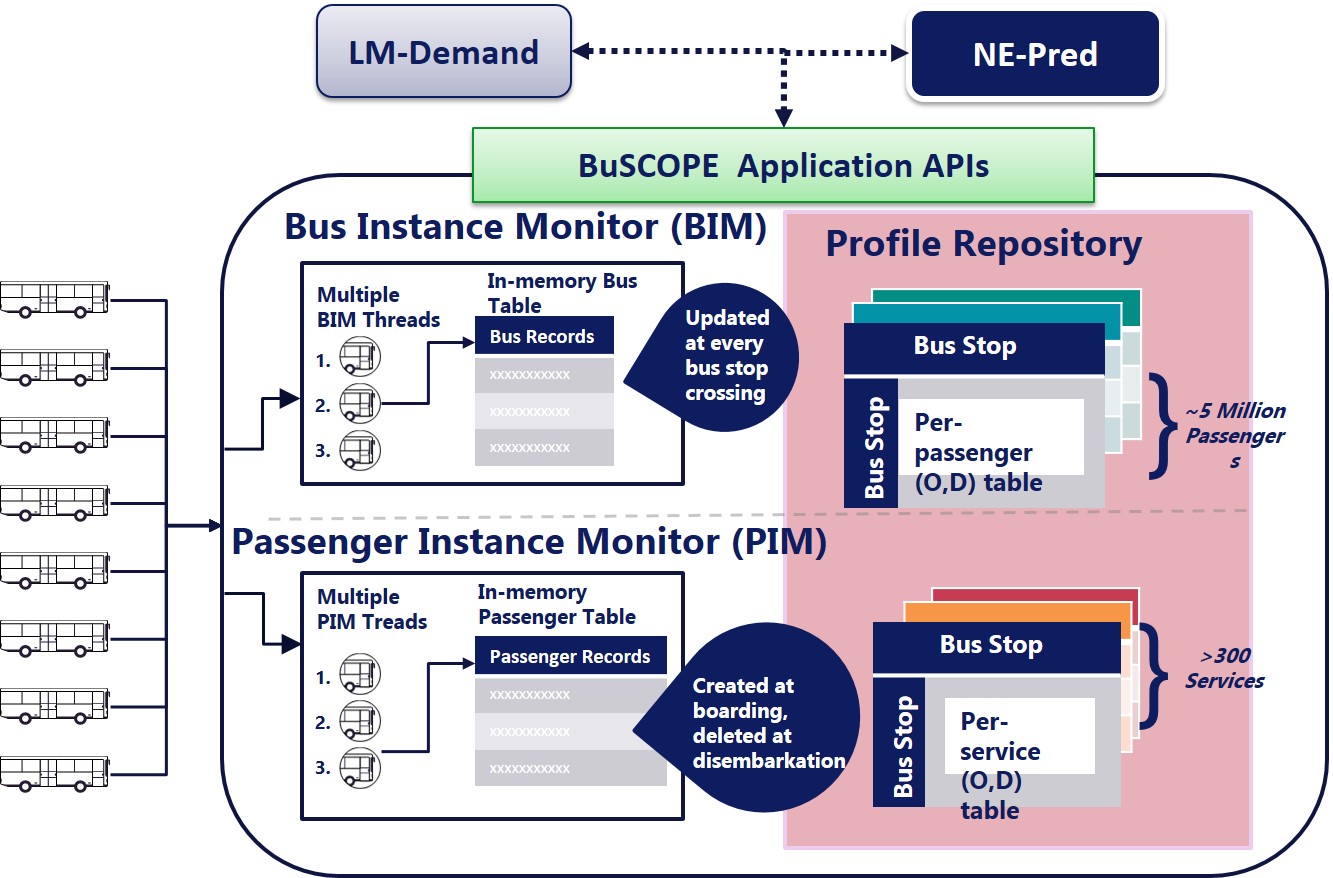}
        \caption{Functional Components of \name}
        \label{fig:systemarch}
    \end{minipage}
\end{center}
\vspace{-0.15in}
\end{figure*}

% \begin{figure}[ht]
%   \includegraphics[width =0.4\textwidth]{pictures/139FlowVisualizationV3.png}
%   \caption{Illustration of Transition (Disembarkation) Probabilities along a typical Bus Service ``139'' in Direction 1}
%   \label{fig:chorddiagramroute130}
% \end{figure}

The results in the previous section were focused on empirically establishing key commuting properties of bus users, thereby motivating the smart city applications that we shall describe later. To support the \emph{live} services that we envision, we now describe our design and implementation of \names, which provides the soft-real time analytics components needed to support multiple smart-city services. As we shall show, the overall number of events of interest across the bus network may appear large (an average of $\approx$ 3.142 million bus commutes each day) but can be supported by a relatively straightforward, multi-threaded, in-memory implementation on a single production-grade server.

Figure~\ref{fig:systemarch} shows the \name middleware architecture, consisting of the following components:
\begin{itemize}[leftmargin=1em]
\item \emph{Bus Event Generator (BEG):} This component resides on the bus and is logically part of its telematics unit. It effectively generates a stream of events, aggregating multiple boarding and disembarkation events into a single payload at each bus-stop, generating an average of 743.40 events/minute (across all buses) during peak hours.
%once every 743.40\kas{} minutes during peak hours).

\item \emph{Passenger Instance Monitor (PIM):} This component logically maintains the state of every passenger currently in transit on any bus in the public transportation network. The incoming data streams from BEG units are de-multiplexed and dispatched to one of multiple PIM threads. The threads operate on a common \emph{In-Memory Passenger Table} (implemented as a hashmap), which maintains a collection of passenger records, indexed by the passenger ID (the $cid$) and the service route. Each record stores, among other fields, a boolean flag indicating whether this is a \emph{regular} passenger or not, and a disembarkation list (with the disembarkation probability for each downstream bus-stop). 

\item \emph{Bus Instance Monitor (BIM):} Analogous to the PIM, this component logically maintains the state of each bus instance that is currently operational. In particular, the incoming data streams from each bus instance is assigned to one of multiple BIM threads, which share a common  hashmap-based \emph{In-Memory Bus Table}. Each record in this table maintains the following bus instance-specific fields: bus location, bus service number, number of on-board passengers, list of on-board passengers (pointers to entries in the In-Memory Passenger Table) and the fraction of on-board passengers classified as \emph{regular}.

\item \emph{Profile Repository:} This component stores the results of the offline analytics that are periodically performed across the entire bus network's transportation data. It computes and stores (a) a passenger-centric profile, which includes a per-passenger, per-service $(O,D)$ matrix (one for each of the 8 day type-time bins described previously) storing the number of past trips for that $(O,D)$ pair; and (b) a  bus-service specific matrix that similarly stores the number of observed $(O,D)$ flows between all bus-stops on the route, aggregated over all passengers.

\end{itemize}

As illustrated in Figure~\ref{fig:systemarch}, the \name system exposes a set of service APIs that are used by the  \emph{LM-Demand} and \emph{NE-Pred} applications.

\subsection{Performance Considerations}
\begin{figure*}[t]
\begin{center}
     \begin{minipage}[t]{0.3\textwidth}
         \includegraphics[width=2in,height=1.5in]{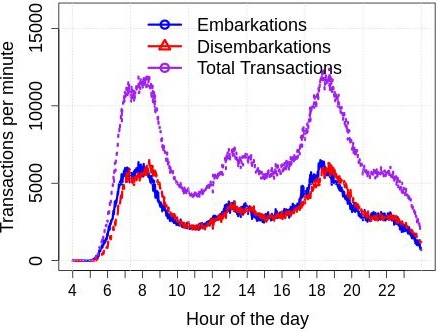}
        \caption{Number of Embarkation/Disembarkation Events}
        \label{fig:eventintensity}
    \end{minipage}
    \hfill
    \begin{minipage}[t]{0.3\textwidth}
        \includegraphics[width=2in,height=1.5in]{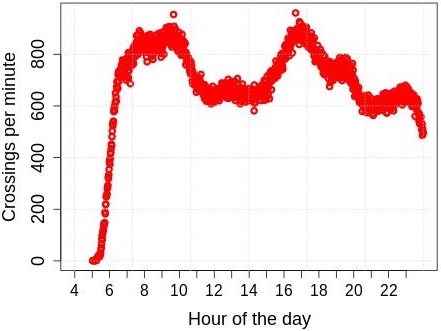}
        \caption{Intensity of Bus-Stop Crossings}
        \label{fig:crossingintensity}
    \end{minipage}
    \hfill
    \begin{minipage}[t]{0.3\textwidth}
        \includegraphics[width=2in,height=1.5in]{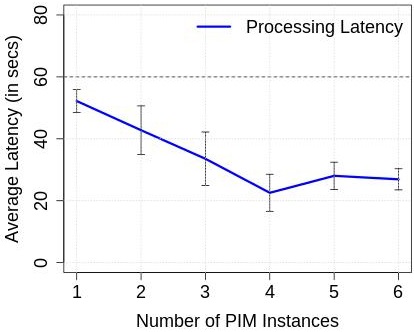}
        \caption{Processing Latency vs. Number of PIM instances}
        \label{fig:throughputevents}
    \end{minipage}
\end{center}
\vspace{-0.15in}
\end{figure*}

To understand the workload characteristics of the \name system and its impact on the system complexity requirements, we first analyzed the historical data to understand (a) the event intensity of  embarkation \& disembarkation events across all buses, and (b) how often we generate a bus-stop crossing event, across the entire city. Figure~\ref{fig:eventintensity} plots the average of the transaction events/min--i.e., the sum of the embarkation and disembarkation events across the entire Singaporean bus network, over different hours of the day, at minute-level granularity. We see that, at peak periods, there are approx. 6000 boarding and alighting events/min on average, with the total reaching approx. 12000 such events--each such event will correspond to an update (either creation or removal) in the corresponding PIM entity. Similarly, Figure~\ref{fig:crossingintensity} plots the number of bus-stop crossings per minute. This metric is relevant as the BIM-specific information needs to be potentially updated only after each crossing (as a bus's state will not change in between bus-stops). We see that, during the rush hour peaks, the maximum rate of generation of such crossing is \~900 crossings/minute, implying that the \name should be able to update one BIM record with an average latency lower than $\sim 60$ msecs.

% \begin{figure*}[t]
% \begin{center}
% \mbox{
% \subfigure[Transaction Rate]{
% \includegraphics[width=3in,height=2in]{pictures/transaction-volume-total.jpg}
%          \caption{Number of Embarkation/Disembarkation Events }
%          \label{fig:eventintensity}
% }
% \subfigure[Crossings Rate]{
% \includegraphics[width=3.5in,height=2in]{pictures/busstopcrossing.jpg}
%         \caption{Intensity of Bus-Stop Crossings}
%         \label{fig:crossingintensity}
% }
% \subfigure[Processing Latency]{
% \includegraphics[scale=0.6]{pictures/pim.jpg}
%     \caption{TBD XXX: Processing Latency/Memory vs. Number of PIM instances}
% \label{fig:throughputevents}
% }
% }
% \caption{Performance Requirements of \name}
% \label{fig:sys-reqs}
% \end{center}
% \end{figure*}

\subsection{\name System Performance}
To satisfy the above performance bounds, the \name implementation, hosted on an Intel Xeon server with 128 GB memory and up to 14 processor, effectively utilizes multiple BIM and PIM threads. Figure~\ref{fig:throughputevents} plots the relationship between the number of PIM components ($N_p$), and the average processing latency, when processing incoming events in per-minute chunks. The experiment is performed using events generated during a 30-minute  evening peak period (6.45 PM to 7.15 PM) from different weekdays. We observed that the memory footprint (consisting primarily of the records of currently on-board passengers throughout the bus network) remains almost invariant at $\approx$ \textbf{10.05 MB}. We can conclude that the use of a modest number (3-4) threads allows \name to comfortably handle even peak event workloads--even a single-threaded implementation takes $\approx$59 seconds to process an entire minute's worth of events, with each individual event incurring an average of 17.33msec processing latency. Similar results are obtained for the BIM components (details omitted due to space constraints): a modest number of threads allows \name to update each bus-specific record with the details of  multiple boarding or alighting passengers at each bus-stop.

%%%% throughput

% \begin{figure}[tbh]
% \begin{center}
% \includegraphics[scale=0.6]{pictures/pim.jpg}
%     \caption{TBD XXX: Processing Latency/Memory vs. Number of PIM instances}
% \label{fig:throughputevents}
% \end{center}
% \end{figure}

% \begin{figure}[tbh]
% \begin{center}
% \includegraphics[scale=0.6]{pictures/placeholder.png}
%     \caption{TBD XXX: Processing Latency/Memory vs. Number of BIM instances}
% \label{fig:buscrossingevents}
% \end{center}
% \end{figure}

%\input{4b.mod.tex}
\section{LM-DEMAND: Predictive Mobility-on-Demand}
\label{sec:lmdemand}
We now tackle the \emph{LM-Demand} application. At a high-level, this application has two components: (a) an analytics component that looks to predict the number of disembarkations at a location (either an individual bus-stops or a collection of nearby bus-stops) sufficiently in advance; and (b) a resource optimization component that uses such prediction values to smartly dispatch and pre-position the MoD vehicles.

Given the commuter dynamics investigated in the previous section, we are now aimed to forecast the disembarkations in a given bus stop by leveraging the personalised and aggregate level commuter traits. More specifically, our tasks are to have the capabilities of (a) user-level disembarkation prediction given that he boarded a bus service from a particular bus stop, and (b) ability to make such predictions with a certain look-ahead time.

\subsection{Hybrid Approach for Disembarkation Prediction}
As we have seen previously, the disembarkation point of \emph{regular} users can be predicted quite accurately at the time of boarding; similarly, the disembarkation point of irregular users can also often be assigned to a limited set of \emph{sink} locations along the route. We propose to build a hybrid model that synthesizes both these approaches, taking advantage of \emph{regularity} (the predictability of travel patterns at an individual level) and \emph{conformity} (the tendency for people to follow flow-based statistics at an aggregate level) to obtain a more precise prediction.

% Figure~\ref{fig:XXX} \am{We need to insert the flow-chart figure here} illustrates the logic of his hybrid approach. 
Our approach works as follows. Based on the entries in the \name repository, each boarding commuter is declared as a \emph{regular} vs. \emph{irregular} passenger for this specific journey. More specifically, consider a commuter $c$ boarding a bus $b$ at bus stop $s$ in a give time-bin (as before, for concreteness, we consider the 4 time-bins corresponding to the tuple (off-peak|peak,  weekend|weekday)). If this boarding pattern for $c$ has \emph{high support} (we'll quantify `high' shortly), then this user is marked as \emph{regular} and $cid$ is projected to disembark at the highest-confidence bus-stop (among the bus-stops remaining in the journey) for this particular boarding pattern. (As mentioned before, the boarding pattern uses the common path re-identification strategy to incorporate the possibility that multiple bus services may provide an equivalent journey for this segment). However, if the user support is low, then $c$ is declared an irregular user, in which case we use the aggregate flow information to declare that $cid$ will disembark at the bus-stop that has the highest flow-based conditional probability from $s$.

%\sx{The $N_{threshold}$ here is a still confusing. $N_{threshold}$ should be a certain value, so i suggest not defining it in relation to $x_i$, as $N_{threshold}$ does not have a $i$ with it. How do u choose $N_{threshold}$? By histogram, or percentile? Any reasoning and logic behind?  In my understanding, $N_{threshold}$ should be set first, irrelevant of $x_i$; instead $x_i$ is derived after $N_{threshold}$ is set. Excuse me if my understanding is wrong.} 

%We use \textit{$x_{threshold}$} \& \textit{$N_{user}$} as two tunable system parameters to determine if a user is regular or not for a specific boarding bus stop: (a) \textit{$N_{threshold}$} represents the fraction of historical trips (relative to the number of trips taken by the most active user, within that (time-window, bus stop) context, in the data set), needed to classify a user as a regular, while (b) \textit{$N_{user}$} is the corresponding user-specific parameter, which is calculated upon a tap-in event based the number of historical trip records of tap-in ($x_{user}$) by the user, at that specific embarkation bus stop. They are defined by:

To formalize this approach, we use \textit{$N_{threshold}$} as a tunable system parameter to determine if a user is regular or not for a specific boarding bus stop. \textit{$N_{threshold}$} represents the fraction of historical trips (relative to the number of trips taken by the most regular user of any single bus stop in the data set) needed to classify a user as a regular:
\begin{small}
\begin{equation}
   N_{threshold}
  = \Big\lfloor\left(\frac{\text{(} x_{threshold} - x_{min}\text{)}}{\text{(}x_{max} - x_{min}\text{)}}100\%\right)\Big\rfloor
\end{equation}

% \begin{equation}
%     N_{user}
%   = \Big\lfloor\left(\frac{\text{(} x_{user} - x_{min}\text{)}}{\text{(}x_{max} - x_{min}\text{)}}100\%\right)\Big\rfloor
% \end{equation}
\end{small}
where $x_{max}$ \& $x_{min}$ ($x_{max}\ge x_{min}\ge 0$) respectively denote the maximum \& minimum number of records (across all users) that originates from \emph{any} single bus stop. (The expression $N_{threshold}$ simply provides a common way to define regularity across different data sets: given $N_{threshold}$, $x_{min}$ and $x_{max}$, we can then compute $x_{threshold}$ ($\in \text{[}x_{min},x_{max}\text{]}$) denoting the actual minimum number of historical trip records which should be present for a traveler to be considered regular). For a given user, boarding with a given (embarkation bus stop, time-window) context, we then compute the user-specific value $x_{user}$, the number of prior embarkations in the data set with the given context, and declare the user's current trip to be \emph{regular} iff $x_{user} \ge x_{threshold}$.

\subsection{Evaluation of Disembarkation Prediction}
To present detailed results, we consider 20 bus services belonging to two distinct classes: 10 buses that pass through the CBD from other areas of Singapore and 10 feeder buses that traverse primarily through NCBD areas. We consider the data of first 3 weeks (comprising 14 weekdays) as a training set and last week of August 2013 as the test set.

% \begin{figure}[!htb]
% \begin{center}
% \mbox{
% \subfigure[CBD Weekday]{
% {\includegraphics[width=1.65 in]{pictures/accuracy_dismbarkation_cbd_weekday.png}}
% \label{fig:dismbarkation_cbd_weekday}
% }
% \subfigure[NCBD Weekday]{
% {\includegraphics[width=1.65 in]{pictures/accuracy_dismbarkation_ncbd_weekday.png}}
% \label{fig:dismbarkation_ncbd_weekday}
% }
% }
% \caption{Accuracy of disembarkation prediction}
% \label{fig:disembarkation-accuracy}
% \end{center}
% \end{figure}
%  kas-guys commenting this out for a bit -- subfigure is causing issues

\begin{figure}[t]
\includegraphics[width=2.4in,height=1.3in]{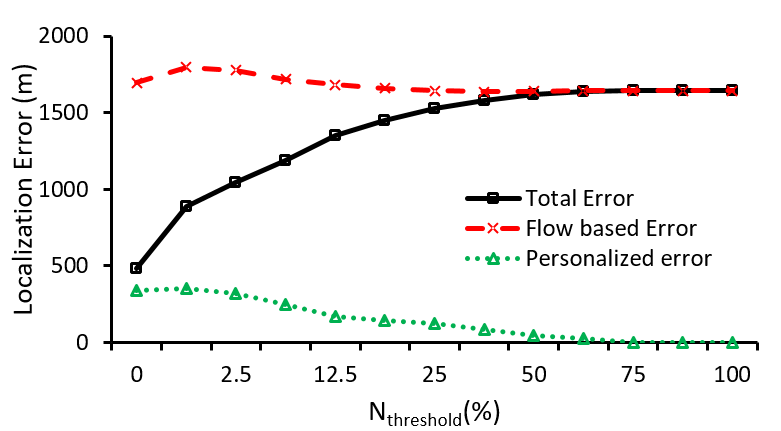}
\caption{CBD Weekday}
\label{fig:disembarkation-accuracy}
\vspace{-0.15in}
\end{figure}

\noindent \textbf{Evaluation metrics:}  We study two key metrics: (a) \emph{Prediction localization error} -- this measure computes the spatial distance between the actual and predicted location of disembarkation; and (b) \emph{Bus ridership estimation} -- this measure computes the error in predicting the number of passengers remaining on-board a bus, and is computed as the difference between the total passengers actually on the bus at a future stop and the number predicted to remain. In addition, we compare our proposed  Hybrid approach against a \emph{Flow-based baseline}~\cite{Zhang18}, where the number of passengers predicted to disembark at bus-stop $d$ (out of the set of passengers $N_b$ who boarded at $b$) is computed as $N_b* P_{bd}$, where $P_{bd}$ is the flow (transition) probability from bus-stop $b$ to $d$. 

Figure~\ref{fig:disembarkation-accuracy} plots the localization error, averaged across all bus instances and all of the 10 routes, separately for buses that travel through CBD on weekdays. The average localization error is plotted as a function of $N_{threshold}$. As $N_{threshold}$ increases, the proportion of trips deemed to be regular diminishes and that of irregular trips increases, as a commuter must have undertaken many more rides to be considered as a candidate for individualized prediction). In particular, for $x_{threshold}=1$, a trip is considered to be regular if there is a history of even one past embarkation by the commuter within that time-bin (i.e., $x_{user} \ge 1$), and along the current route. We plot both the total average error, as well as the errors for the regular and irregular trips separately. We see that at the left-most extreme point (i.e., all trips with non-zero support value classified as regular) provides the least localization error, of approx. 500 meters (~1-2 bus stops). In contrast, the Flow-based baseline conforms to the right most extreme point (i.e., all users classified as non-regular) and incurs an error of >1.5 km (>6 bus stops). Note that, as expected, as $N_{threshold}$ increases, the average personalized error decreases as the set of regular trips are now restricted to only those that are observed even more dominantly and thus represent highly predictable commutes (e.g., home-to-office). However, the fraction of trips deemed regular also decreases, and the higher contribution of flow-based errors increases the overall error rate; hence, we use $x_{threshold}=1$ for our subsequent analyses.

\begin{figure}[t]
\includegraphics[width=2.4in,height=1.5in]{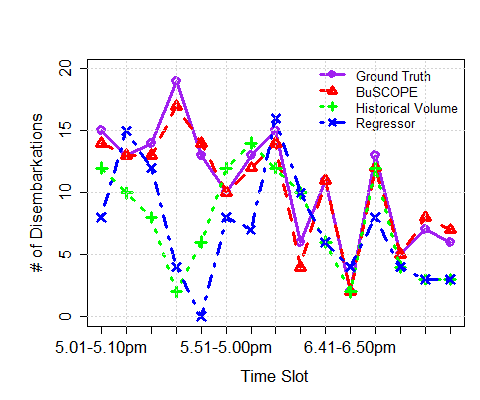}
\caption{BuScope vs. Baseline (historical) predictors}
\label{fig:benchmarks}
\vspace{-0.1in}
\end{figure}

\begin{table}[ht]
\begin{small}
\centering
\begin{tabular}{c c c}
   \hline 
   &    CBD (in meters) & NCBD (in meters) \\ \hline \hline 
   Weekday &  480.3  & 342.8  \\ 
   Weekend &  548.4  & 558.1  \\ \hline \hline 
\end{tabular}
\caption{Localization error of Disembarkation Prediction}
\label{tab:disembarkaccuracy}
\end{small}
\vspace{-0.1in}
\end{table}

In addition to the error, we also studied the accuracy of disembarkation prediction--i.e., the fraction of trips an exact prediction of the alighting stop was made. From Table~\ref{tab:disembarkaccuracy}, which tabulates this accuracy for all 4 spatiotemporal bins, we see that the accuracy is generally above 85\%.

Furthermore, we compare our hybrid approach \name with 2 new baseline strategies that are both based on \emph{aggregated} analysis of historical commuting data:
\begin{itemize}[leftmargin=1em]

    \item[(a)] \emph{Historical Volume:} This approach computes (and uses as the predicted value), for each bus stop, the average number of aggregate disembarkations observed historically within a specific temporal (e.g., time-of-day, day-of-week) window.
    
    \item[(b)] \emph{Regressor:} This approach constructs a linear regression model (per bus stop) with the following covariates: time-of-day, day-of-week and the number of buses seen to historically transit through that bus stop within that time window. This regression model is then used to estimate the disembarkation; note that this model is not predictive as it needs the retrospectively reconstructed ground-truth of the number of transiting buses.
\end{itemize}
In Figure~\ref{fig:benchmarks}, we plot the number of predicted disembarkations for a 10-minute time window (depicted in X-axis over a 2.5-hour period) at a city-hub bus stop that serves an urban campus, local museums, and various businesses. We observe that \name tallies with the ground truth (actual disembarkations) much better, achieving 92.59\% accuracy; in contrast the \emph{Historical Volume} and \emph{Regressor} strategies achieve only 55.56\% and 62.96\% accuracy, respectively. By performing similar analysis over the entire set of bus stops and bus routes, we find that \name achieves a significant (over 30\%) accuracy improvement over both baselines.

\textbf{Dynamic Disembarkation Prediction and Lookahead Time:} The results above require the prediction of a trip's disembarkation bus-stop right at the point of boarding. In a slightly more sophisticated, dynamic version, the disembarkation predicted is updated dynamically, as the journey progresses. In particular, if the passenger remains on-board when the bus passes the currently predicted disembarkation stop, the prediction is updated to the downstream stop with the highest conditional probability. Accordingly, one can anticipate that the prediction accuracy increases as the journey progresses and the bus gets nearer to the true stop. We empirically found that this dynamic prediction accuracy was significantly higher (89\% accuracy) when the prediction was made ~13 mins (corresponding to ~9 bus stops) in advance of the actual disembarkation\footnote{This lookahead time vs. accuracy tradeoff will be used in our analysis of predictive MoD placement.}.

\begin{figure*}[t]
\begin{center}
     \begin{minipage}[t]{0.3\textwidth}
         \includegraphics[width=2.1in,height=1.2in]{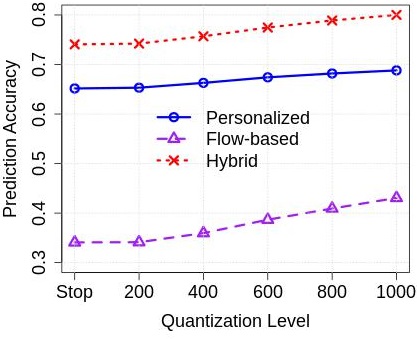}
         \caption{Impact of Spatial Quantization on Prediction Accuracy}
         \label{fig:accuracyvsquantization}
    \end{minipage}
    \hfill
    \begin{minipage}[t]{0.3\textwidth}
        \includegraphics[width=2.1in,height=1.2in]{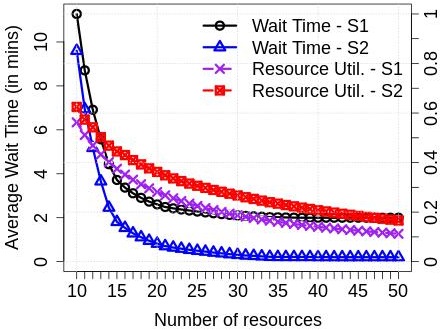}
        \caption{Waiting Time vs. Resource Utilization with $T_D$=2mins ($C$=1)}
        \label{fig:mod-toapayoh-2mins}
    \end{minipage}
    \hfill
    \begin{minipage}[t]{0.3\textwidth}
        \includegraphics[width=2.1in,height=1.2in]{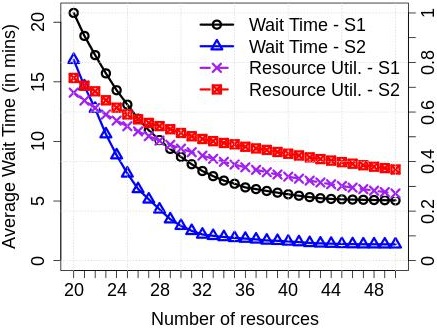}
        \caption{Waiting Time vs. Resource Utilization with $T_D$=5mins ($C$=1)}
        \label{fig:mod-toapayoh-5mins}
    \end{minipage}
\end{center}
\vspace{-0.15in}
\end{figure*}

\begin{figure}[t]
\includegraphics[width=2.4in]{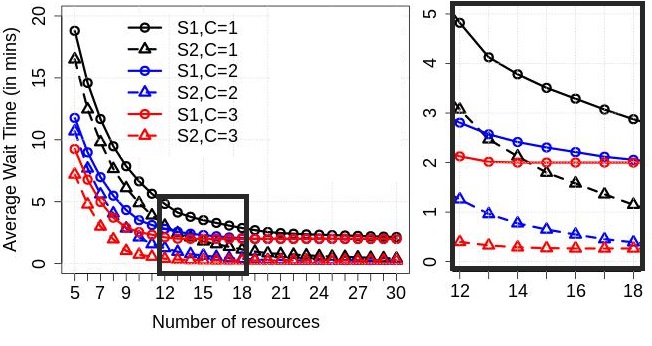}
\caption{Impact on Waiting Times with Varying Capacity, $C$, of MoD Vehicles (left), with zoomed-in view for 12 to 18 vehicles (right).}
\label{fig:vary_capacity}
% \am{Can we add in a zoomed `inset' for say (x=12-18)? so that the reviewer can see the minor differences?}
\vspace{-0.15in}
\end{figure}

\textbf{Impact of Spatial Granularity on Prediction Accuracy:} We also investigated the impact of the spatial granularity on the prediction accuracy level. We first mapped the individual bus stops island-wide to grids of size $N_{g} = (200, 400, 600, 800, 1000)$ meters, and re-ran the predictions. Figure~\ref{fig:accuracyvsquantization} plots the accuracy for different grid sizes, across \textbf{all} bus services (not just the 20 previously mentioned), for a typical weekday, AM peak period. As anticipated, the accuracy of prediction improves in all three cases (i.e., personalized, flow-based and hybrid), reaching 80\% for $N_g=800$ meters.

\subsection{Predictive MoD Performance}
\label{sec:mod-perf}
Having established the ability to predict disembarkation locations with high accuracy, we now show that such predictions can be used to improve the performance of last-mile MoD systems. We consider the case of a specific Singapore neighborhood\footnote{https://data.gov.sg/dataset/master-plan-2014-subzone-boundary-web} (Toa Payoh, one of Singapore's central residential estates) and focus on the disembarkation behavior of passengers across a representative region consisting of 16 bus-stops. We specifically consider a two hour window around the PM-Peak period (when a large number of commuters may be expected to return home)--this region sees, on average, $\approx$ 300 disembarkations during this period.

Given that a last-mile MoD system does not currently exist, we utilize a simulation framework to model the MoD system. We make a few simplifying assumptions: (a) each vehicle's capacity is $C$ passengers and passengers are serviced in First-come-First-served (FCFS) fashion; (b) the final destination of a last-mile passenger is randomly distributed within the region, and is modeled by a constant travel time (from the bus-stop) of $T_D$ mins; (c) similarly, unless a vehicle is at the disembarking bus-stop, it will take $T_D$ mins to arrive there from its current location. In addition, we assume the availability of accurate travel time estimates (now widely available via various applications in cities such as Singapore) and thus assume that the arrival time at a bus-stop is known via external means.  

We study two different strategies:
\begin{itemize}[leftmargin=1em]
    \item \emph{Reactive MoD Strategy (\textbf{S1})}: Under this strategy, we assume that a vehicle remains stationary after dropping its current complement of passengers, and moves to the next bus-stop whenever there is a waiting passenger there (thereby causing the passenger to experience a wait time of \emph{at least} $T_D$ mins). Of course, if the vehicle is currently busy, it must first complete its current set of dropoffs, before heading back for the next passenger.
    \item \emph{Proactive MoD Strategy (\textbf{S2})}: Under this strategy, if a vehicle is free, and a set of disembarkations are predicted to happen in the future, the vehicle proactively moves to the bus-stop with the earliest such predicted disembarkation. If the next disembarkation actually occurs more than $T_D$ mins after the vehicle became free, then the passenger will experience zero wait; else, his wait time will be the difference between the vehicle's arrival time and the true disembarkation time. Moreover, because a vehicle takes at most $2T_D$ minutes to respond, we perform disembarkation prediction of bus passengers with a look-ahead time, $T_l = 2T_D$ mins. To handle possible incorrect predictions where a vehicle being allocated to \emph{anticipated} passengers who don't show up eventually, we set an appropriate \emph{expiry} parameter to free up the resource.
\end{itemize}

Figures~\ref{fig:mod-toapayoh-2mins} and ~\ref{fig:mod-toapayoh-5mins} plot the average wait time (across all disembarking passengers) for the two strategies as a function of the number of MoD vehicles, for the simplest case $C=1$. We simulated two cases: (a) with $T_D = 5$mins and (b) $T_D = 2$mins (the latter one corresponds to a realistic last-mile travel distance of ~300 meters, assuming an MoD speed of 10km/h). We see that our proactive approach (S2) results in short wait times---an average of $<30$secs for $T_D=2$, and $<2$mins for $T_D$=5, when the number of MoD vehicles is sufficient ($\ge 30$). More importantly, this wait time is around 75\% lower than that experienced by the baseline Reactive approach (S1).

In practice, we expect MoD vehicles to be \emph{shared} by a number of passengers, and not be allocated solely for a single passenger per trip. In Figure~\ref{fig:vary_capacity}, we plot the average wait times for both strategies with $T_D = 2$mins, and the capacity ($C$) of the vehicle being varied from 1 to 3. We simplify the assignment task by clumping together the $C-$closest passengers arriving, or expected to arrive, in a greedy manner. As observed previously, the proactive approach provides a significant reduction in wait times across all values of $C$; as expected, higher values of $C$ achieve comparable wait times with fewer vehicles. For instance, with $C = 3$, the number of resources required reduces by two-thirds (from 30 to 10) for a comparable 75\% reduction in wait time. As the assignments happen every epoch (i.e., a minute in our case), it is possible that the last vehicle to be assigned during an epoch to be not filled to its maximum capacity -- in other words, if the number of remaining unassigned passengers is less than $C$, the passengers are assigned an available vehicle right away, instead of being delayed till the next epoch.

\section{Urban Event Anomaly Detection} \label{sec:anomaly}
We now detail \emph{NE-Pred}, which uses the \name system (which provides live tracking of regular vs. irregular passengers on each operating bus) to enable detection of urban events. At a high level,
our approach is to first identify commuter-level anomalies (irregular trip patterns--i.e., those with zero prior support) for each bus, assign an anomaly score to each bus \& bus-stop based on such anomalies, and finally perform spatial aggregation across all the buses \& multiple bus-stops to identify and localize such events. Our proposed approach has three components:

\begin{figure*}[thb]
\begin{center}
     \begin{minipage}[t]{0.3\textwidth}
          \includegraphics[width=2.3in]{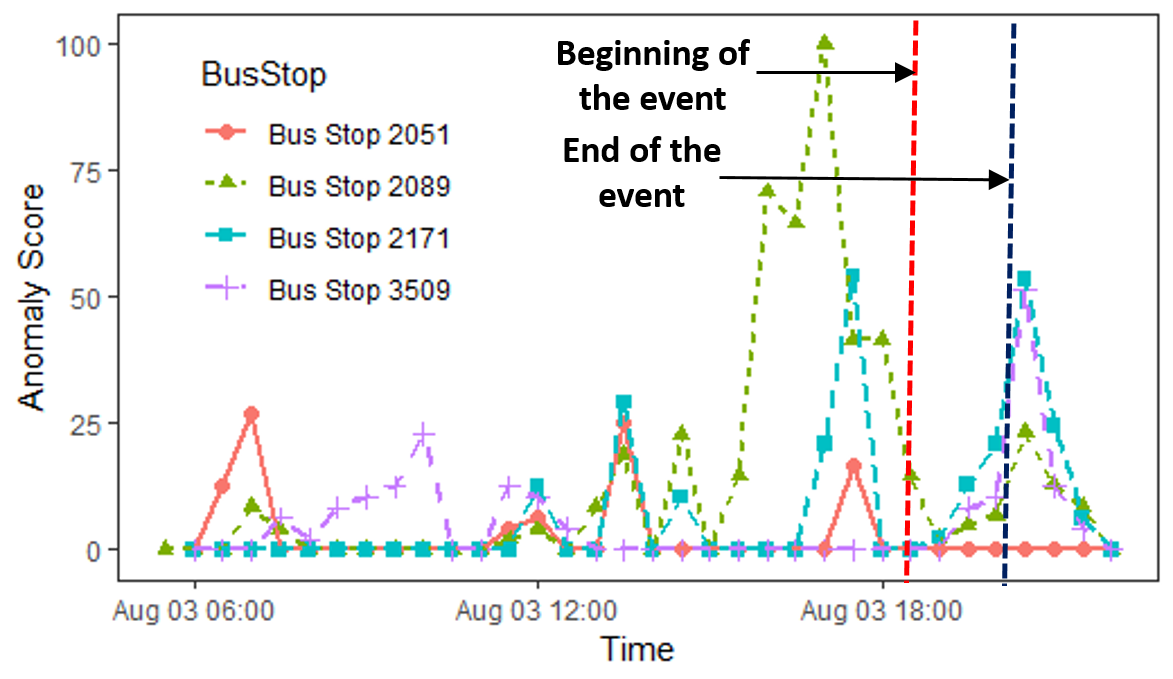}
          \caption{Temporal variation of $Anom(.)$ at nearby bus stops (NDR)}
          \label{fig:temporalanomaly_ndpr}
     \end{minipage}
    \hfill
    % \begin{minipage}[t]{0.3\textwidth}
    %     \includegraphics[width=2.3in]{pictures/AnomalyScoresAug24SOTA_v2.png}
    %     \caption{Temporal variation of $Anom(.)$ at nearby bus-stops (SCH)}
    %      \label{fig:temporalanomaly_sota}
    % \end{minipage}
    % \hfill
    \begin{minipage}[t]{0.3\textwidth}
      \includegraphics[width=2.0in,height=1.5in]{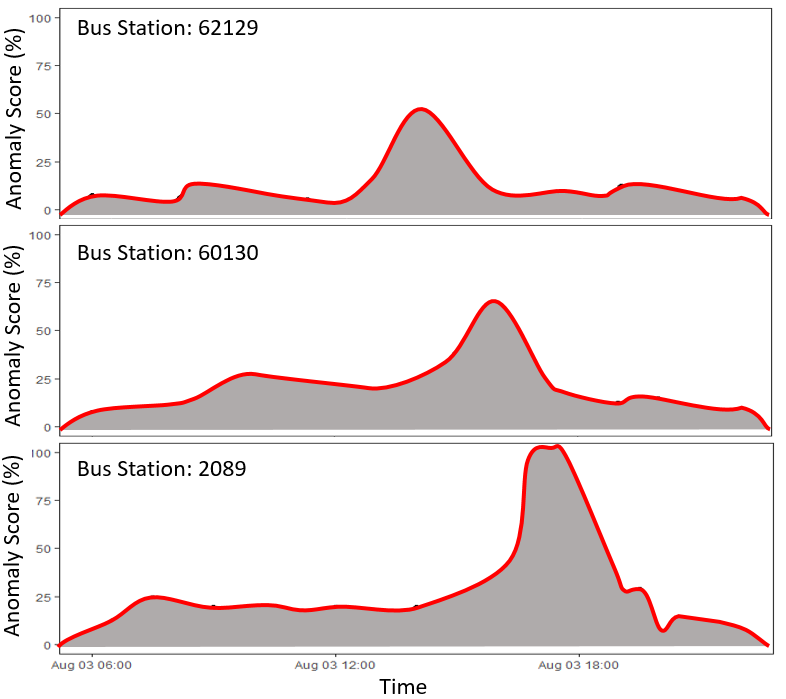}
      \caption{Spatial Evolution of Anomaly Score (NDR)}
      \label{fig:anomalyevolution}
    \end{minipage}
    \hfill
    \begin{minipage}[t]{0.3\textwidth}
      \includegraphics[width=2.3in]{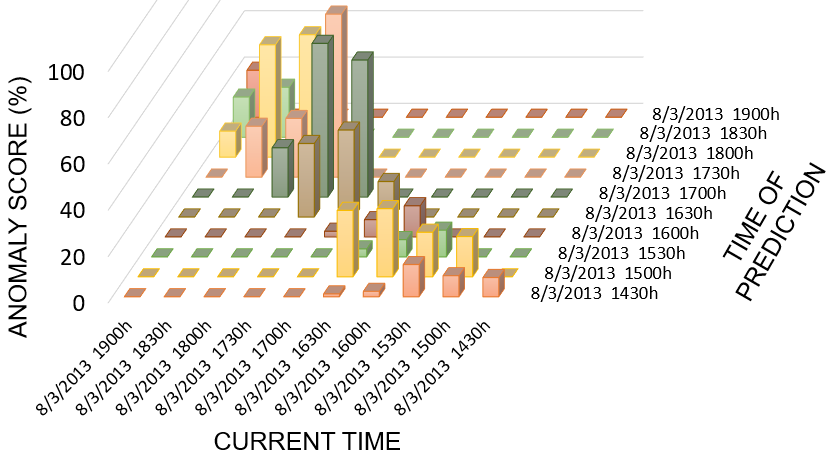}
      \caption{Predictive forecasting (bus 36, bus stop 2051) on 08/03/13}
      \label{fig:predictivescore}
    \end{minipage}
\end{center}
\vspace{-0.15in}
\end{figure*}

\begin{itemize}[leftmargin=1em]
\item \textbf{Event Detection:} We show that bus-stops near an event venue see a sharp spike in the volume of either boarding or disembarkation by irregular users, \emph{across multiple bus services that transit through those bus-stops}, and thus derive an anomaly threshold-based strategy to detect such events. 

\item \textbf{Event Localization:} We then derive a 2-D agglomerative clustering approach (that computes a cluster representative in space domain) to identify the location of such detected events. By using the \emph{cascade anomaly} we detect the onset of anomaly by propagating the bus-stop and epoch specific anomaly scores to the downstream bus stops. The goal here is to minimize the spatial error and detect the events well in advance.

\item \textbf{Event Prediction:} Finally, by using a novel \emph{time-shifted anomaly extrapolation} technique, a variant of cascade anomaly described earlier, where bus-specific anomalies are \emph{temporally} extrapolated to future downstream bus-stops, we show that we can identify the start-time and location of such events well in advance (well before visitors begin to show anomalous disembarkation patterns \emph{at} the event venue).
\end{itemize}

\subsection{Computing Anomaly Scores}
Both event detection and localization rely on a fundamental technique: computing the anomaly score/contribution from a bus that has just visited a specific bus-stop. Intuitively, at a given bus-stop, if the bus sees either passenger disembarkations (visitors heading to an event) or boardings (perhaps residents avoiding an event) that are \emph{irregular}, its anomaly score will be higher and indicative of an event in the vicinity of that bus-stop. This anomaly score is updated for each bus and assigned to the corresponding bus-stop after each such bus-stop crossing. To mathematically express the computed anomaly, for a given bus $b$, at stop $s$, let us denote by $o_i$ the interaction of an \emph{irregular individual} (i.e., an individual with zero support) $i$. (For simplicity we will drop indices $b$ and $s$; the score is computed in the same way for each bus at each bus-stop.) The interaction value $o_i=1$ if the irregular individual exhibits one of two possible interactions with $b$ at $s$,  $ =\{\text{embark,disembark} \}$ (action set denoted by $ =\{e,d\}$ for short). On the other hand, $o_i=0$ if $i$ does not interact with the bus at the stop, i.e. $i$ is not observed at the bus-stop or $i$ simply remains on the bus, having boarded earlier and heading to a different stop. 

The definition of an irregular interaction is driven by the support of the corresponding events (embarkation or disembarkation) at $s$ for the corresponding context (i.e., the AM/PM \& peak/off-peak time-bins). Of course, additional contextual states (e.g., the weather) may affect such commuting patterns and should ideally be incorporated for our specific dataset, note that climatic conditions are relatively stable across Singapore in August (a fairly dry month).

%which is defined to include: (a) the day-of-week $dow \in \{ \text{weekday}, \text{weekend} \}$;  and (b) time-of-day $tod \in [t-\delta, t+\delta]$. Note that, in contrast to our past time-bins defined in terms of coarser AM/PM \& peak/off-peak periods, we define the temporal context at finer granularity (we typically use $\delta=XXX$ mins) for \emph{NE-Pred} as our goal is to identify the start of events with O(mins) accuracy. For notational ease, we will simply refer to the set of context attributes as $C=\{dow,tod \in [t-\Delta, t+\Delta],s\}$. 

The anomaly count at a given bus stop $s$ for a bus instance $b$ is simply given as number of non-regular commuters who interact with $b$ (i.e., board or alight) at $s$. The nominal anomaly score $A(s,t)$, for the bus-stop $s$ is then computed, in units of time $\Delta$ (=30 mins), by aggregating the anomaly count of all buses that traverse bus stop $s$ during the time $[t-\Delta, t+\Delta]$. The final anomaly score of a stop $s$ at time $t$, denoted by $Anom(s,t)$ is obtained as the bus-stop specific, \emph{normalized} deviation of the current score, i.e., $Anom(s,t)= \frac{A(s,t)- min_{\forall \tau}\{A(s,\tau)\}}{max_{\forall \tau}\{A(\tau)\}-min_{\forall \tau}\{A(\tau)\}}$.

\subsection{Experimental Results}
To quantitatively evaluate our event anomaly detection algorithm, we consider a relatively small set of events (tabulated in Table~3) whose occurrence during August 2013 were well documented\footnote{While we were able to scrape many other events, obtaining reliable estimates of the start times of such past events proved very difficult. We believe that these 3 events are adequate for demonstrating our approach.}. 

\begin{table}[t]
\begin{small}
\begin{center}
    \resizebox{1\linewidth}{!}{
        \begin{tabular}{p{2.4cm}>{\arraybackslash}m{1.5cm} >{\arraybackslash}m{1.7cm} >{\arraybackslash}m{1.1cm} >{\arraybackslash}m{.7cm} >{\arraybackslash}m{0.5cm}} 
        \hline
        Event               & Location & Date \& Time & Spatial Error (m) & \multicolumn{2}{c}{Look-ahead time (mins)} \\
                            &          &              &               & Cascade & Spot  \\ 
        \hline \hline 
        National day        & Float @      & 3-Aug           & 345.21 & 60  & 20  \\
        rehearsal (NDR)     & Marina bay   & 06:30-08.30PM &        &           \\ \hline
        National day        & Float @      & 9-Aug           & 376.51 & 210 & 130 \\
        parade (NDP)        & Marina bay   & 06:30-08:30PM &        &           \\ \hline
        Franz Schubert      & SOTA         & 24-Aug          & 669.71 & 30 & NA   \\
        Piano sonatas (SCH) & Concert Hall & 07:30-08:30PM &        &           \\ 
        \hline \hline 
        \end{tabular}
    }
\end{center}
\end{small}
\caption{Summary of Events and Localization Results}
\label{tab:events}
\vspace{-0.4 in}
\end{table}

\subsubsection{Detecting Urban Event Anomalies}
% and ~\ref{fig:temporalanomaly_sota}
Figures~\ref{fig:temporalanomaly_ndpr}  plots the temporal variation of anomaly scores of neighboring bus-stops for the NDR event day. We see that $Anom(s)$ is appreciably higher at those stops, close to the event start (as well as end) times. (In fact, while further discrimination between event start vs. end is possible by differentiating between $e$ and $d$ interactions, we omit this discussion for space reasons.) In general, we observe that a rule ``$Anom(s) \ge 50\% \; $for two consecutive intervals $t_i,t_{i+1}$'' helps us to accurately identify all 3 representative events.

% \begin{figure}[!thb]
% \begin{center}
% \includegraphics[width=3.2 in]{pictures/Prediction3DDiagram_v2.png}
% \vspace{-0.1in}
% \caption{Predictive score forecasting for bus service 36 at bus stop number 2051 on 3rd of August 2013}
% \vspace{-0.1in}
% \label{fig:predictivescore}
% \end{center}
% \end{figure}

\subsubsection{Event Anomaly Localization}
We now describe our clustering-based strategy for spatial event localization. For each bus-stop $s$, let $t_p(s)$ denote the time at which $Anom(s)$ peaks, while $l_s$ represents the 2-D location of $s$. We employ a greedy hierarchical agglomerative technique for spatial event localization. Intuitively, we start by \emph{merging} the two bus-stops with the larger $Anom(s)$ values into a single cluster, after which we iteratively pick the bus-stop (among the set of bus-stops remaining to be clustered) with the highest value of $Anom(s)$ and merge it with our cluster. The merging operation involves computing the weighted centroid of the location ($l_s$) and the current cluster.

To identify the start time of an anomaly, we adopt a cascading technique where $Anom(s,t)$ of a specific bus stop $s$ and bus $b$ at time $t$ is propagated to all its downstream bus stops and re-assigned at each bus stop-crossing. We continually update this anomaly score (for each bus stop) at each successive epoch (with $\Delta$ = 30 mins); an anomaly is then declared to have ``started", when a bus stop's score, $Anom(s,t)$ exceeds the threshold for 2 consecutive epochs.

Table~\ref{tab:events} shows the resulting spatial error and the look-ahead time for all 3 events. We see that the spatial error is around 350-400 meters (roughly $\approx$ 1.5 bus-stops) for the larger-scale National Day events. The slightly higher error for $SCH$ may be explained by noting that the event was approx. 200 meters distant from a major station, where many visitors probably disembarked and then walked to the venue. We also note that the cascading technique yields greater look-ahead time (varying between 60 and 210 minutes) for macro events (NDR and NDP) as compared to the micro events (SCH), most likely because, at large events, visitor arrivals peak well before the event start--e.g., on the NDP day, hordes of visitors arrive at least 3 hours ahead to obtain favorable viewing spots.

To further demonstrate the advantages of the cascading approach, we also introduce a naive baseline strategy called \emph{Spot Anomaly}, where an anomaly is defined for each bus stop in isolation (based on changes solely in that bus stop's disembarkation volume. In Table~\ref{tab:events}, we report the look-ahead time of this baseline -- clearly, our \emph{Cascade} method detects the macro events well in advance (40-80 minutes before) of the baseline. Note also that, unlike \emph{Spot}, \emph{Cascade} was able to detect the micro event (SCH). As explained earlier, this limitation may be attributed to the fact that $SCH$ was located near a major transit hub. In such a case, the overall change in disembarkation volume at a busy `sink' node is likely to be insignificant, causing \emph{Spot} to fail. However, \emph{Cascade's} technique of using abnormal occupancy on multiple bus routes can isolate such low-intensity events.

%\am{There is some logical flaw here. Table 3 is supposed to represent, the temporal ``look-ahead time"--i.e., how far in advance do we declare that any anomaly will happen. Lookahead time is different that temporal prediction accuracy. E.g., 2 hrs in adance, I tell you that event will start at 6pm. It starts at 6.15pm. In this case, lookahead=2hr 15 min, whereas temporal error is 15 min. From the table, I don't see how we make the claim that "SCH prediction is more precise"--to me, it seems that the table is merely saying that ``I can detect SCH only 30 min in advance". We should be clear about this reasoning. In fact, as I read further--i.e., I read Section 6.2.3, I'm now a bit alarmed. It seems that the `lookahead time' computation is only in Section 6.2.3 and plotted in Figure 20. If so, then the entries called `temporal margin' in Table 3 are actually 'temporal prediction error'. If so, the results suggest that Spot has lower error (prediction closer to the actual start time) than Cascade. Our writing and logic would need to change quite significantly if this were true.}

% \begin{table}[ht]
% \centering
% \begin{tabular}{|c||c |c|}
%   \hline
%   &    Spatial Error (meters) & Temporal Margin (mins)  \\ \hline \hline
%   NDR. &  345.21  & 60  \\ \hline
%   NDP  &  376.51  & 210   \\ \hline
%   SCH & 669.71 & 30  \\ \hline
% \end{tabular}
% \caption{Accuracy of Anomaly Detection}
% \label{tab:eventlocaccuracy}
% \end{table}

\subsubsection{Event Prediction}
The results above show that we can in fact use the implicit signals from bus commuting patterns to detect an event's location and look-ahead time with high accuracy. However, we now show that we can achieve something even more powerful: we can predict the start time of an unknown event \emph{well in advance}. The key idea is as follows: bus passengers travelling to participate in an event will often board the buses well in advance (e.g., the average commute from Singapore's residential heartland to the downtown area is over 45 minutes). By effectively propagating such anomalous boarding signals to downstream bus-stops, we can identify the possible future start-time and location of such events. More specifically, our algorithm operates as follows:
\begin{itemize}[leftmargin=1em]
    \item Compute the anomaly score $Anom(b,s,t)$ for a given bus $b$ that traverses bus-stop $s$ at current time $t$.
    \item Based on the estimated travel time (denoted as $T(s,\hat{s})$ of bus $b$ to a downstream stop $\hat{s}$), propagate this anomaly score to $\hat{s}$ for  future time-instant--i.e., let \\ $PredAnom(b,\hat{s},t+ T(s,\hat{s}))$= $Anom(b,s,t)$.
    \item For each downstream bus-stop $\hat{s}$, aggregate anomaly scores across all buses that will travel to $\hat{s}$.
    \item If the predicted anomaly score at any bus-stop $\hat{s}$ exceeds the threshold at any future time $t+T$, then declare ``event likely at $\hat{s}$ at time $t+T$".
\end{itemize}

Figure~\ref{fig:predictivescore} illustrates this concept of predictive anomaly score propagation, using a specific bus-service (No. 36) on the day of the NDR event. We can see that the predicted anomaly score exceeds our threshold (50\%) for two consecutive periods at 5pm, and identifies the event start-time as 5.30pm. In other words, we are able to \emph{correctly predict the occurrence of the event 1.5 hours in advance}. Similar results hold for the other events, demonstrating the promise of our proposed method. 

%\begin{figure}[!thb]
%\begin{center}
%\includegraphics[width=3.2 in]{pictures/SpatialEvolutionAnomalyScore.png}
%\vspace{-0.1in}
%\caption{Spatial Spread of Anomaly Score at 1730h on 2013-08-09}
%\vspace{-0.1in}
%\label{fig:anomalyevolution}
%\end{center}
%\end{figure}

%\begin{figure}[!thb]
%\begin{center}
%\includegraphics[width=3.2 in]{pictures/DistantBusStops.PNG}
%\vspace{-0.1in}
%\caption{Spatial Evolution of Anomaly Score on 2013-08-03}
%\vspace{-0.1in}
%\label{fig:anomalyevolution}
%\end{center}
%\end{figure}

%\begin{figure}[!thb]
%\begin{center}
%\includegraphics[width=3.2 in]{pictures/LocalizationError.PNG}
%\vspace{-0.1in}
%\caption{Spatial localization of the NDP Rehersals}
%%\vspace{-0.1in}
%\label{fig:predictivelocalization}
%\end{center}
%\end{figure}

\section{Discussion} \label{sec:discussion}
There are several aspects of live bus ridership analytics that need additional investigations.

\noindent \textbf{Threats to Validity:} As mentioned previously, it is possible for passengers to pay their fare by cash to the driver, in which case they are essentially invisible to our analysis. While relatively rare (only 4\% of trips\footnote{https://www.researchgate.net/publication/266878969\_use\_of\_public\_transport\_smart\_card\\ \_fare\_payment\_data\_for\_travel\_behavior\_analysis\_in\_singapore} involve cash), certain groups of commuters (e.g., overseas tourists on short trips) may favor such transactions. Our analytical results may consequently be less accurate for locations disproportionately favored by tourists. Also, the predictions on commuter demand patterns should ideally made more holistically, factoring in other modes of transport (e.g., the train network, private on-demand buses, etc.). 
%It is possible that some of the commuting characteristics observed here (such as the existence of 2-3 sink nodes on each route) may not hold for these other transportation modes.

%Nonetheless, we believe that our inferences are robust as (i) the \emph{LM-Demand} application is particularly relevant for suburban residential neighborhoods where tourist traffic should be much lower, and (ii) the preference of tourists to use cash for their ostensibly \emph{irregular} trips makes \emph{NE-Pred} less susceptible to spurious event detection. O

\begin{figure}[!thb]
\begin{center}
\includegraphics[width=0.40\textwidth]{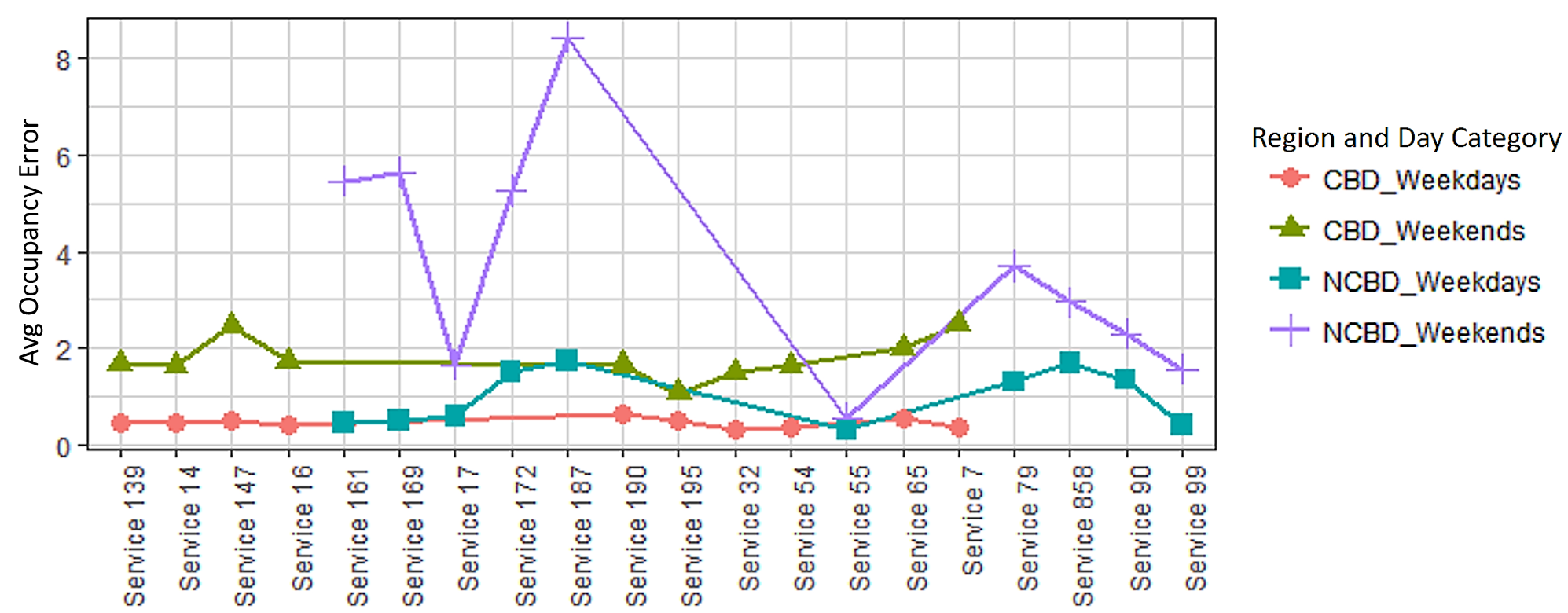}
\caption{Accuracy of ridership estimation}
\label{fig:ridership}
\vspace{-0.15in}
\end{center}
\end{figure}

\noindent \textbf{Other Application Scenarios:} Live, predictive disembarkation prediction can enable other types of smart transportation services. For example, commuters often use transportation Apps\footnote{https://busleh.originallyus.sg/;\\https://play.google.com/store/apps/details?id=com.iridianstudio.sgbuses\&hl=enSG} that provide "live" feeds of bus arrival times and crowdedness. By using disembarkation prediction, such Apps can provide a commuter, waiting at a particular bus-stop, a more accurate, \emph{anticipated} crowdedness of an en-route bus, as opposed to simply displaying the current crowd levels. Figure~\ref{fig:ridership} plots the average error in occupancy prediction using our Hybrid prediction technique--we see that the average error in predicting ridership at downstream bus-stops is almost always quite low (<2 persons), and may be thus used to enhance such transportation Apps.

%This type of predictive estimates might enable a commuter to realize, for example, that an apparently crowded bus that is 3 stops away may is very likely to become empty at the previous bus-stop (a ``sink" node). 

\noindent \textbf{Smarter MoD Allocation Strategies:} We must emphasize that the benefit of lower last-mile wait times illustrated here has been done using a fairly straightforward MoD simulator model. Significant opportunities for optimizing the MoD resource allocation exist--for example, the vehicle assignment may be dynamically updated based on the real last-mile travel distances. Our goal here was not to present a preferred strategy, but simply to empirically demonstrate that disembarkation predication can significantly improve the last-mile commuting experience.

% \sx{To make the last-mile MoD solution more sustainable, the number of autonomous vehicles is also important factor to consider and thus the allocation strategy might minimize the vehicle fleet size while incorporating waiting time policy to guarantee the quality of service [refA], or (d) The MoD resource allocation may be further extended with first-mile demand to make it holistic solution for real-world deployment [refA]S. P. Chuah, S. Xiang and H. Wu, "Optimal rebalancing with waiting time constraints for a fleet of connected autonomous taxi," 2018 IEEE 4th World Forum on Internet of Things (WF-IoT), Singapore, 2018, pp. 629-634.}
%\textbf{Generalizability:} \kas{other cities with differing characteristics, commuter volume etc}

% \begin{figure}
%   \includegraphics[keepaspectratio, width=9cm]{pictures/DatasetContribution.PNG}
%   \caption{Dataset Distribution Based on Card Types}
%   \label{fig:datasetcontribution}
% \end{figure}

\noindent \textbf{Data Privacy:} The use of even pseudonymized data (as we do) can raise possible privacy concerns, such as the possible recovery of a user's identity from detailed individual level mobility traces--e.g., a daily pattern of early morning embarkations and  evening alightings would identify the ``home" bus stop for a pseudo-identifier. Clearly, there is a risk of privacy compromise by possibly cross-linking such inferences to publicly side available information (e.g.,~\cite{Zang2011}). To get an initial sense of this problem, we conducted a preliminary assessment of the $k-anonymity$ of a typical `terminal' bus stop--i.e., we ask: ``on average, how many unique customers would have the same bus stop as their \emph{`home'}"? For a specific neighborhood, we first extracted the locations of residential blocks and used a \emph{Nearest Neighbor classifier} to assign them to a set of predefined clusters (bus stops are the centroids). By then estimating the total population within each cluster and multiplying it by the bus ridership ratio ($\approx$0.32)\footnote{\url{https://data.gov.sg/group/transport}}, we find that the k-anonymity values can range from 24 -- 392 (for low housing density areas such as Marine Parade) to around 79 -- 1500 (for more mature residential neighborhoods, such as Toa Payoh). These results suggest that bus stop-level data may typically not be trivial to de-anonymize--however, more careful assessment of privacy vs. data granularity (and its impact on our analytics) remains an area for further research.

\section{Related Work} \label{sec:relatedwork}
The widespread availability of city-scale mobility data (obtained via GPS/WiFi~\cite{Zhou16,  loecher2009citysense} traces, taxi ridership~\cite{Chiang15} or bike trip~\cite{Zhang18_2} records, public transport data~\cite{Yuan13,jayarajah2018}) have driven significant prior research on urban mobility analytics.

\noindent \textbf{Human Mobility Prediction}
Prior work has shown that human macro-scale mobility is regular and predictable in both spatial and temporal instances~\cite{Kim07, Wang11, Wang15, Oliveira16}. Regular and frequent visiting patterns (e.g., home, work and supermarket) enables the ability to predict human mobility with high accuracy~\cite{Eagle09, Li10}. In particular, in~\cite{Gonzalez08} the authors show the existence of individual-level frequent and routine visits to a few locations. In~\cite{Song10}, it is shown that the human mobility is accurately predictable on college campuses (93\% accuracy). Works such as those of Becker et al.\cite{becker2013human} utilize metropolitan-scale mobility data (from CDR records) to characterize population movement for use cases such as commute time predictions and disease spreading. Besides characterizing such predictability, researchers have also worked on predicting future locations based on historical mobility traces. In recent past Markov models are proposed to predict the future location~\cite{Kotz:2004, realitymining, Xue13, Xue15, Krumm06} in both indoors and outdoors~\cite{gambs2012next} by utilising historical transition between places. The authors in~\cite{gomes2013will} complemented the historical traces by leveraging various contextual information inferred by exploiting various sensors (bluetooth, accelerometer etc). In this work, we leverage the existence of `regularity' in human mobility patterns (observed only sporadically through public transport usage) at both individual and collective scales.

\noindent \textbf{Demand Prediction of Urban Transport Networks}
Prior works focused on analyzing the demand of public transport by modelling spatiotemporal historical demand and availability of vehicles~\cite{Li08, Chang10, Chiang15}. Similarly, demand estimation in ride sharing/ride hailing (i.e., MOD) has garnered significant attention, specially after the emergence of services such as Uber. In such environments, the focus was often on predicting the arrival rate of the passengers at a given location to re-position the vehicles to cater future demand~\cite{Pavone12, Zhang16, Davis16}. Closest to our work, Balan et al.\cite{balan2011real} provide real-time trip information services based on historic trips (e.g., fare and distance estimates of similar trips in the past), and discuss use cases such as anomaly detection -- our work is different in that we focus on soft-real time guarantees but operate on live, city-scale streaming mobility data. 

\noindent \textbf{Urban Event Detection} Work here can be classified in sub-domains of urban event detection and prediction and anomaly detection in transport (particularly in road networks). For event detection, works such as CitySense~\cite{loecher2009citysense} utilise aggregated GPS traces collected using a mobile application to detect hot-spots and anomalies/outbreaks. This approach is similar to our \emph{Spot} anomaly baseline which looks at aggregate disembarkations to detect outliers. Konishi et al.\cite{Konishi:2016} have recently proposed an approach to \emph{predict} irregularities (e.g., large scale events) ahead of time using a two-step modeling process. By querying route information using a mobile transit App, the authors model short and long term population models using auto-regression and bi-linear Poisson regression, respectively. Similarly, social media data has been used to detect and track earthquakes from user posted information on Twitter~\cite{Sakaki:2010} and to detect and characterize urban events from text, images and metadata~\cite{jayarajah2016can}.

%using a transit App used by mobile users where they are able to query route information tuple (origin location, destination location, a time in future),

Previous works on transportation anomaly detection have looked at varied aspects such as detection of anomalies, understanding the spatiotemporal ordering and finding root causes. Pang et al. \cite{Pang:2011} detect contiguous, spatiotemporal cells as anomalous regions using Likelihood Ratio Tests. Further, Liu et al. \cite{Liu:2011} proposed a formulation for ``causal outlier detection" for detecting the emergence, propagation and disappearance of outliers (e.g., traffic jam). Subsequently, Chawla et al. \cite{Chawla:2012} identified routes in a road network with anomalous traffic using a 2-step approach:  (1) first they detect anomalous links using Principal Component Analysis (as seen in many works on network traffic anomaly detection) and (2) using a link-route matrix, they detect which routes were root causes for the detected anomalies using L1 machinery. In contrary, our work aims to detect and localize events by specifically exploiting the  inherent `regularity' of \emph{individual-level} human mobility.
%They represent the road network and flows between them as a graph of regions and detect spatio-temporal using the minimum distortion measure. More importantly, they infer the causal relationship of the detected anomalies using a frequent subtree algorithm that is loosely based on association rule mining
%Regularity
% 3. Lin M, Hsu WJ, Lee ZQ. Predictability of Individuals’ Mobility with High-resolution Positioning Data. In: Proceedings of the 2012 ACM Conference on Ubiquitous Computing. ACM; 2012. p. 381–390.
% 4. Smith G, Wieser R, Goulding J, Barrack D. A Refined Limit on the Predictability of Human Mobility. In: Pervasive Computing and Communications (PerCom), 2014 IEEE International Conference on. IEEE
\section{Conclusions} \label{sec:conclusion}
We have described \names, a system that supports soft-real time processing of public bus commuting data. From the analysis of such individualized bus trip records, we have shown that the destination for most trips has high predictability, even on routes where the commuter has made only one past journey. Subsequently, by combining individualized and flow-based predictions, we show that we can predict a commuter's disembarkation bus-stop with an accuracy of over 85\% and a mean error of less than 1-2 bus-stops. We have then shown how such collective predictions can be used in a last-mile MoD system, where unmanned vehicles are pre-positioned to respond to anticipate disembarkation demand, resulting in an $\approx$75\% decrease in commuter wait times. We have also shown how the real-time detection of irregular commuters, along multiple bus routes, can be used to detect urban events with high spatial accuracy ($\approx$450 meter error), well in advance (100 mins) of the event start time. We anticipate that this work will motivate public agencies to view mobility data as not just a policy planning resource, but as an enabler of a new class of \emph{live} smart city services.

% \section{Ridership Prediction} \label{sec:ridership}
% \input{5.Ridership.tex}

\section{Acknowledgment}
This material is supported partially by the National Research Foundation, Prime Minister's Office, Singapore under its International Research Centres in Singapore Funding Initiative and under NRF-NSFC Joint Research Grant Call on Data Science (NRF2016NRF-NSFC001-113), and partially by the Air Force Research Laboratory, under agreement number FA2386-14-1-002. K. Jayarajah's work was supported by an A*STAR Graduate Scholarship. The view and conclusions contained herein are those of the authors and should not be interpreted as necessarily representing the official policies or endorsements, either expressed or implied, of the Air Force Research Laboratory or the US Government.

\bibliographystyle{ACM-Reference-Format}
\balance
\bibliography{11.references}


\begin{thebibliography}{00}

%%% the Web address.

\ifx \showCODEN    \undefined \def \showCODEN     #1{\unskip}     \fi
\ifx \showDOI      \undefined \def \showDOI       #1{#1}\fi
\ifx \showISBNx    \undefined \def \showISBNx     #1{\unskip}     \fi
\ifx \showISBNxiii \undefined \def \showISBNxiii  #1{\unskip}     \fi
\ifx \showISSN     \undefined \def \showISSN      #1{\unskip}     \fi
\ifx \showLCCN     \undefined \def \showLCCN      #1{\unskip}     \fi
\ifx \shownote     \undefined \def \shownote      #1{#1}          \fi
\ifx \showarticletitle \undefined \def \showarticletitle #1{#1}   \fi
\ifx \showURL      \undefined \def \showURL       {\relax}        \fi
% invisible to TeX
\providecommand\bibfield[2]{#2}
\providecommand\bibinfo[2]{#2}
\providecommand\natexlab[1]{#1}
\providecommand\showeprint[2][]{arXiv:#2}

\bibitem[\protect\citeauthoryear{Balan, Nguyen, and Jiang}{Balan
  et~al\mbox{.}}{2011}]%
        {balan2011real}
\bibfield{author}{\bibinfo{person}{Rajesh~Krishna Balan},
  \bibinfo{person}{Khoa~Xuan Nguyen}, {and} \bibinfo{person}{Lingxiao Jiang}.}
  \bibinfo{year}{2011}\natexlab{}.
\newblock \showarticletitle{Real-time trip information service for a large taxi
  fleet}. In \bibinfo{booktitle}{{\em Proceedings of the 9th international
  conference on Mobile systems, applications, and services}}. ACM,
  \bibinfo{pages}{99--112}.
\newblock


\bibitem[\protect\citeauthoryear{Becker, C{\'a}ceres, Hanson, Isaacman, Loh,
  Martonosi, Rowland, Urbanek, Varshavsky, and Volinsky}{Becker
  et~al\mbox{.}}{2013}]%
        {becker2013human}
\bibfield{author}{\bibinfo{person}{Richard Becker}, \bibinfo{person}{Ram{\'o}n
  C{\'a}ceres}, \bibinfo{person}{Karrie Hanson}, \bibinfo{person}{Sibren
  Isaacman}, \bibinfo{person}{Ji~Meng Loh}, \bibinfo{person}{Margaret
  Martonosi}, \bibinfo{person}{James Rowland}, \bibinfo{person}{Simon Urbanek},
  \bibinfo{person}{Alexander Varshavsky}, {and} \bibinfo{person}{Chris
  Volinsky}.} \bibinfo{year}{2013}\natexlab{}.
\newblock \showarticletitle{Human mobility characterization from cellular
  network data}.
\newblock \bibinfo{journal}{{\it Commun. ACM}} \bibinfo{volume}{56},
  \bibinfo{number}{1} (\bibinfo{year}{2013}), \bibinfo{pages}{74--82}.
\newblock


\bibitem[\protect\citeauthoryear{Castro, Zhang, and Li}{Castro
  et~al\mbox{.}}{2012}]%
        {castro2012urban}
\bibfield{author}{\bibinfo{person}{Pablo~Samuel Castro},
  \bibinfo{person}{Daqing Zhang}, {and} \bibinfo{person}{Shijian Li}.}
  \bibinfo{year}{2012}\natexlab{}.
\newblock \showarticletitle{Urban traffic modelling and prediction using large
  scale taxi GPS traces}. In \bibinfo{booktitle}{{\em International Conference
  on Pervasive Computing}}. Springer, \bibinfo{pages}{57--72}.
\newblock


\bibitem[\protect\citeauthoryear{Chang, Tai, and Hsu}{Chang
  et~al\mbox{.}}{2010}]%
        {Chang10}
\bibfield{author}{\bibinfo{person}{H. Chang}, \bibinfo{person}{Y. Tai}, {and}
  \bibinfo{person}{J.~Y. Hsu}.} \bibinfo{year}{2010}\natexlab{}.
\newblock \showarticletitle{{Context-aware Taxi Demand Hotspots Prediction}}.
\newblock \bibinfo{journal}{{\em Business Intelligence and Data Mining\/}}
  (\bibinfo{year}{2010}).
\newblock


\bibitem[\protect\citeauthoryear{Chawla, Zheng, and Hu}{Chawla
  et~al\mbox{.}}{[n. d.]}]%
        {Chawla:2012}
\bibfield{author}{\bibinfo{person}{Sanjay Chawla}, \bibinfo{person}{Yu Zheng},
  {and} \bibinfo{person}{Jiafeng Hu}.} \bibinfo{year}{[n. d.]}\natexlab{}.
\newblock \showarticletitle{Inferring the Root Cause in Road Traffic
  Anomalies}. In \bibinfo{booktitle}{{\em Proceedings of the 2012 IEEE 12th
  International Conference on Data Mining}} {\em (\bibinfo{series}{ICDM '12})}.
\newblock


\bibitem[\protect\citeauthoryear{Chiang, Hoang, and Lim}{Chiang
  et~al\mbox{.}}{2015}]%
        {Chiang15}
\bibfield{author}{\bibinfo{person}{M.~F. Chiang}, \bibinfo{person}{T.~A.
  Hoang}, {and} \bibinfo{person}{E.~P. Lim}.} \bibinfo{year}{2015}\natexlab{}.
\newblock \showarticletitle{{Where are the Passengers? A Grid-based Gaussian
  Mixture Model for Taxi Bookings}}. In \bibinfo{booktitle}{{\em ACM
  International Conference on Advances in Geographic Information Systems
  (SIGSPATIAL)}}.
\newblock


\bibitem[\protect\citeauthoryear{Colpaert, Chua, Verborgh, Mannens, Van~de
  Walle, and Vande~Moere}{Colpaert et~al\mbox{.}}{2016}]%
        {colpaert2016public}
\bibfield{author}{\bibinfo{person}{Pieter Colpaert}, \bibinfo{person}{Alvin
  Chua}, \bibinfo{person}{Ruben Verborgh}, \bibinfo{person}{Erik Mannens},
  \bibinfo{person}{Rik Van~de Walle}, {and} \bibinfo{person}{Andrew
  Vande~Moere}.} \bibinfo{year}{2016}\natexlab{}.
\newblock \showarticletitle{What public transit API logs tell us about travel
  flows}. In \bibinfo{booktitle}{{\em Proceedings of the 25th International
  Conference Companion on World Wide Web}}. International World Wide Web
  Conferences Steering Committee, \bibinfo{pages}{873--878}.
\newblock


\bibitem[\protect\citeauthoryear{Davis, Raina, and Jagannathan}{Davis
  et~al\mbox{.}}{2016}]%
        {Davis16}
\bibfield{author}{\bibinfo{person}{N. Davis}, \bibinfo{person}{G. Raina}, {and}
  \bibinfo{person}{K. Jagannathan}.} \bibinfo{year}{2016}\natexlab{}.
\newblock \showarticletitle{{A Multi-level Clustering Approach for Forecasting
  Taxi travel Demand}}. In \bibinfo{booktitle}{{\em Intelligent Transportation
  Systems (ITSC)}}.
\newblock


\bibitem[\protect\citeauthoryear{Eagle and Pentland}{Eagle and
  Pentland}{2006}]%
        {realitymining}
\bibfield{author}{\bibinfo{person}{Nathan Eagle} {and} \bibinfo{person}{Alex
  Pentland}.} \bibinfo{year}{2006}\natexlab{}.
\newblock \showarticletitle{Reality mining: sensing complex social systems}.
\newblock \bibinfo{journal}{{\em Personal and ubiquitous computing\/}}
  \bibinfo{volume}{10}, \bibinfo{number}{4} (\bibinfo{year}{2006}),
  \bibinfo{pages}{255--268}.
\newblock


\bibitem[\protect\citeauthoryear{Eagle and Pentland}{Eagle and
  Pentland}{2009}]%
        {Eagle09}
\bibfield{author}{\bibinfo{person}{N. Eagle} {and} \bibinfo{person}{A.~S.
  Pentland}.} \bibinfo{year}{2009}\natexlab{}.
\newblock \showarticletitle{{EigenBehaviors: Identifying Structure in
  Routine}}.
\newblock \bibinfo{journal}{{\em Behavioral Ecology and Socio-biology\/}}
  (\bibinfo{year}{2009}).
\newblock


\bibitem[\protect\citeauthoryear{Gambs, Killijian, and del Prado~Cortez}{Gambs
  et~al\mbox{.}}{2012}]%
        {gambs2012next}
\bibfield{author}{\bibinfo{person}{S{\'e}bastien Gambs},
  \bibinfo{person}{Marc-Olivier Killijian}, {and}
  \bibinfo{person}{Miguel~N{\'u}{\~n}ez del Prado~Cortez}.}
  \bibinfo{year}{2012}\natexlab{}.
\newblock \showarticletitle{Next place prediction using mobility markov
  chains}. In \bibinfo{booktitle}{{\em Proceedings of the First Workshop on
  Measurement, Privacy, and Mobility}}. ACM, \bibinfo{pages}{3}.
\newblock


\bibitem[\protect\citeauthoryear{Gomes, Phua, and Krishnaswamy}{Gomes
  et~al\mbox{.}}{2013}]%
        {gomes2013will}
\bibfield{author}{\bibinfo{person}{Jo{\~a}o~B{\'a}rtolo Gomes},
  \bibinfo{person}{Clifton Phua}, {and} \bibinfo{person}{Shonali
  Krishnaswamy}.} \bibinfo{year}{2013}\natexlab{}.
\newblock \showarticletitle{Where will you go? mobile data mining for next
  place prediction}. In \bibinfo{booktitle}{{\em International Conference on
  Data Warehousing and Knowledge Discovery}}. Springer,
  \bibinfo{pages}{146--158}.
\newblock


\bibitem[\protect\citeauthoryear{Gonzalez, Hidalgo, and Barbasi}{Gonzalez
  et~al\mbox{.}}{2008}]%
        {Gonzalez08}
\bibfield{author}{\bibinfo{person}{M.~C. Gonzalez}, \bibinfo{person}{C.~A.
  Hidalgo}, {and} \bibinfo{person}{A.~L. Barbasi}.}
  \bibinfo{year}{2008}\natexlab{}.
\newblock \showarticletitle{{Understanding Individual Human Mobility
  Patterns}}.
\newblock \bibinfo{journal}{{\em Nature\/}} (\bibinfo{year}{2008}).
\newblock


\bibitem[\protect\citeauthoryear{Jayarajah and Misra}{Jayarajah and
  Misra}{2016}]%
        {jayarajah2016can}
\bibfield{author}{\bibinfo{person}{Kasthuri Jayarajah} {and}
  \bibinfo{person}{Archan Misra}.} \bibinfo{year}{2016}\natexlab{}.
\newblock \showarticletitle{Can Instagram posts help characterize urban
  micro-events?}. In \bibinfo{booktitle}{{\em Information Fusion (FUSION), 2016
  19th International Conference on}}. IEEE, \bibinfo{pages}{130--137}.
\newblock


\bibitem[\protect\citeauthoryear{Jayarajah, Subbaraju, Athaide, Meegahapola,
  Tan, and Misra}{Jayarajah et~al\mbox{.}}{2018}]%
        {jayarajah2018}
\bibfield{author}{\bibinfo{person}{Kasthuri Jayarajah},
  \bibinfo{person}{Vigneshwaran Subbaraju}, \bibinfo{person}{Noel Athaide},
  \bibinfo{person}{Lakmal Meegahapola}, \bibinfo{person}{Andrew Tan}, {and}
  \bibinfo{person}{Archan Misra}.} \bibinfo{year}{2018}\natexlab{}.
\newblock \showarticletitle{Can Multimodal Sensing Detect and Localize
  Transient Events?}. In \bibinfo{booktitle}{{\em Proceedings Volume 10635,
  Ground/Air Multisensor Interoperability, Integration, and Networking for
  Persistent ISR IX}} {\em (\bibinfo{series}{SPIE Defense + Security '18})}.
\newblock


\bibitem[\protect\citeauthoryear{Kim and Kotz}{Kim and Kotz}{2007}]%
        {Kim07}
\bibfield{author}{\bibinfo{person}{M. Kim} {and} \bibinfo{person}{D. Kotz}.}
  \bibinfo{year}{2007}\natexlab{}.
\newblock \showarticletitle{{Periodic Properties of User Mobility and
  Access-point Popularity}}.
\newblock \bibinfo{journal}{{\em Pervasive and Mobile Computing\/}}
  (\bibinfo{year}{2007}).
\newblock


\bibitem[\protect\citeauthoryear{Konishi, Maruyama, Tsubouchi, and
  Shimosaka}{Konishi et~al\mbox{.}}{2016}]%
        {Konishi:2016}
\bibfield{author}{\bibinfo{person}{Tatsuya Konishi}, \bibinfo{person}{Mikiya
  Maruyama}, \bibinfo{person}{Kota Tsubouchi}, {and} \bibinfo{person}{Masamichi
  Shimosaka}.} \bibinfo{year}{2016}\natexlab{}.
\newblock \showarticletitle{CityProphet: City-scale Irregularity Prediction
  Using Transit App Logs}. In \bibinfo{booktitle}{{\em Proceedings of the 2016
  ACM International Joint Conference on Pervasive and Ubiquitous Computing}}
  {\em (\bibinfo{series}{UbiComp '16})}.
\newblock


\bibitem[\protect\citeauthoryear{Krumm and Horovitz}{Krumm and
  Horovitz}{2006}]%
        {Krumm06}
\bibfield{author}{\bibinfo{person}{J. Krumm} {and} \bibinfo{person}{E.
  Horovitz}.} \bibinfo{year}{2006}\natexlab{}.
\newblock \showarticletitle{{Predestination: Infferring destinations from
  partial trajectories}}. In \bibinfo{booktitle}{{\em ACM International Joint
  Conference on Pervasive and Ubiquitous Computing (Ubicomp)}}.
\newblock


\bibitem[\protect\citeauthoryear{Li, Shin, and Park}{Li et~al\mbox{.}}{2008}]%
        {Li08}
\bibfield{author}{\bibinfo{person}{J. Li}, \bibinfo{person}{I. Shin}, {and}
  \bibinfo{person}{G.~L. Park}.} \bibinfo{year}{2008}\natexlab{}.
\newblock \showarticletitle{{Analysis of Passenger Pick-up Pattern for Taxi
  Location Recommendation}}. In \bibinfo{booktitle}{{\em NCM}}.
\newblock


\bibitem[\protect\citeauthoryear{Li, Ding, Han, Kays, and Nye}{Li
  et~al\mbox{.}}{2010}]%
        {Li10}
\bibfield{author}{\bibinfo{person}{Z. Li}, \bibinfo{person}{B. Ding},
  \bibinfo{person}{J. Han}, \bibinfo{person}{R. Kays}, {and}
  \bibinfo{person}{P. Nye}.} \bibinfo{year}{2010}\natexlab{}.
\newblock \showarticletitle{{Mining Periodic Behaviors for Moving Objects}}. In
  \bibinfo{booktitle}{{\em ACM SIGKDD}}.
\newblock


\bibitem[\protect\citeauthoryear{Liu, Hou, Biderman, Ratti, and Chen}{Liu
  et~al\mbox{.}}{2009}]%
        {Liu2009}
\bibfield{author}{\bibinfo{person}{Liang Liu}, \bibinfo{person}{Anyang Hou},
  \bibinfo{person}{Assaf Biderman}, \bibinfo{person}{Carlo Ratti}, {and}
  \bibinfo{person}{Jun Chen}.} \bibinfo{year}{2009}\natexlab{}.
\newblock \showarticletitle{Understanding individual and collective mobility
  patterns from smart card records: A case study in Shenzhen}.
\newblock \bibinfo{journal}{{\em 12th International IEEE Conference on
  Intelligent Transportation Systems (ITS)\/}} (\bibinfo{year}{2009}),
  \bibinfo{pages}{1--6}.
\newblock


\bibitem[\protect\citeauthoryear{Liu, Zheng, Chawla, Yuan, and Xing}{Liu
  et~al\mbox{.}}{[n. d.]}]%
        {Liu:2011}
\bibfield{author}{\bibinfo{person}{Wei Liu}, \bibinfo{person}{Yu Zheng},
  \bibinfo{person}{Sanjay Chawla}, \bibinfo{person}{Jing Yuan}, {and}
  \bibinfo{person}{Xie Xing}.} \bibinfo{year}{[n. d.]}\natexlab{}.
\newblock \showarticletitle{Discovering Spatio-temporal Causal Interactions in
  Traffic Data Streams}. In \bibinfo{booktitle}{{\em Proceedings of the 17th
  ACM SIGKDD International Conference on Knowledge Discovery and Data Mining}}
  {\em (\bibinfo{series}{KDD '11})}.
\newblock


\bibitem[\protect\citeauthoryear{Loecher and Jebara}{Loecher and
  Jebara}{2009}]%
        {loecher2009citysense}
\bibfield{author}{\bibinfo{person}{Markus Loecher} {and} \bibinfo{person}{Tony
  Jebara}.} \bibinfo{year}{2009}\natexlab{}.
\newblock \showarticletitle{CitySense: Multiscale space time clustering of gps
  points and trajectories}. In \bibinfo{booktitle}{{\em Proceedings of the
  Joint Statistical Meeting}}.
\newblock


\bibitem[\protect\citeauthoryear{Oliveira, Viana, Sarraute, Brea, and
  Hamelin}{Oliveira et~al\mbox{.}}{2015}]%
        {Oliveira16}
\bibfield{author}{\bibinfo{person}{E.~M.~R. Oliveira}, \bibinfo{person}{A.~C.
  Viana}, \bibinfo{person}{C. Sarraute}, \bibinfo{person}{J. Brea}, {and}
  \bibinfo{person}{I.~A. Hamelin}.} \bibinfo{year}{2015}\natexlab{}.
\newblock \showarticletitle{{On The Regularity of Human Mobility}}.
\newblock \bibinfo{journal}{{\em Pervasive and Mobile Computing\/}}
  (\bibinfo{year}{2015}).
\newblock


\bibitem[\protect\citeauthoryear{Pan, Qi, Wu, Zhang, and Li}{Pan
  et~al\mbox{.}}{2013}]%
        {Pan2013}
\bibfield{author}{\bibinfo{person}{Gang Pan}, \bibinfo{person}{Guande Qi},
  \bibinfo{person}{Zhaohui Wu}, \bibinfo{person}{Daqing Zhang}, {and}
  \bibinfo{person}{Shijian Li}.} \bibinfo{year}{2013}\natexlab{}.
\newblock \showarticletitle{Land-Use Classification Using Taxi GPS Traces}.
\newblock \bibinfo{journal}{{\em IEEE Transactions on Intelligent
  Transportation Systems\/}}  \bibinfo{volume}{14} (\bibinfo{year}{2013}),
  \bibinfo{pages}{113--123}.
\newblock


\bibitem[\protect\citeauthoryear{Pang, Chawla, Liu, and Zheng}{Pang
  et~al\mbox{.}}{[n. d.]}]%
        {Pang:2011}
\bibfield{author}{\bibinfo{person}{Linsey~Xiaolin Pang},
  \bibinfo{person}{Sanjay Chawla}, \bibinfo{person}{Wei Liu}, {and}
  \bibinfo{person}{Yu Zheng}.} \bibinfo{year}{[n. d.]}\natexlab{}.
\newblock \showarticletitle{On Mining Anomalous Patterns in Road Traffic
  Streams}. In \bibinfo{booktitle}{{\em Proceedings of the 7th International
  Conference on Advanced Data Mining and Applications - Volume Part II}} {\em
  (\bibinfo{series}{ADMA'11})}.
\newblock


\bibitem[\protect\citeauthoryear{Pavone, Smith, Frazzoli, and Rus}{Pavone
  et~al\mbox{.}}{2012}]%
        {Pavone12}
\bibfield{author}{\bibinfo{person}{M. Pavone}, \bibinfo{person}{S.~L. Smith},
  \bibinfo{person}{E. Frazzoli}, {and} \bibinfo{person}{D. Rus}.}
  \bibinfo{year}{2012}\natexlab{}.
\newblock \showarticletitle{{Robotic Load balancing for Mobility On-demand
  Systems}}.
\newblock \bibinfo{journal}{{\em Robotics Research\/}} (\bibinfo{year}{2012}).
\newblock


\bibitem[\protect\citeauthoryear{Sakaki, Okazaki, and Matsuo}{Sakaki
  et~al\mbox{.}}{[n. d.]}]%
        {Sakaki:2010}
\bibfield{author}{\bibinfo{person}{T. Sakaki}, \bibinfo{person}{M. Okazaki},
  {and} \bibinfo{person}{Y. Matsuo}.} \bibinfo{year}{[n. d.]}\natexlab{}.
\newblock \showarticletitle{Earthquake Shakes Twitter Users: Real-time Event
  Detection by Social Sensors}. In \bibinfo{booktitle}{{\em Proceedings of the
  19th International Conference on World Wide Web}} {\em (\bibinfo{series}{WWW
  '10})}.
\newblock


\bibitem[\protect\citeauthoryear{Song, Qu, Blumm, and Barbasi}{Song
  et~al\mbox{.}}{2010}]%
        {Song10}
\bibfield{author}{\bibinfo{person}{C. Song}, \bibinfo{person}{Z. Qu},
  \bibinfo{person}{N. Blumm}, {and} \bibinfo{person}{A.~L. Barbasi}.}
  \bibinfo{year}{2010}\natexlab{}.
\newblock \showarticletitle{{Limits of Predictability in Human Mobility}}.
\newblock \bibinfo{journal}{{\em Science\/}} (\bibinfo{year}{2010}).
\newblock


\bibitem[\protect\citeauthoryear{Song, Kotz, Jain, and He}{Song
  et~al\mbox{.}}{2003}]%
        {Kotz:2004}
\bibfield{author}{\bibinfo{person}{Libo Song}, \bibinfo{person}{David Kotz},
  \bibinfo{person}{Ravi Jain}, {and} \bibinfo{person}{Xiaoning He}.}
  \bibinfo{year}{2003}\natexlab{}.
\newblock \showarticletitle{Evaluating Location Predictors with Extensive Wi-Fi
  Mobility Data}.
\newblock \bibinfo{journal}{{\em SIGMOBILE Mob. Comput. Commun. Rev.\/}}
  \bibinfo{volume}{7}, \bibinfo{number}{4} (\bibinfo{date}{Oct.}
  \bibinfo{year}{2003}), \bibinfo{pages}{64--65}.
\newblock
\showISSN{1559-1662}


\bibitem[\protect\citeauthoryear{Stephan, Martin, and Theo}{Stephan
  et~al\mbox{.}}{2004}]%
        {Krygsman2004}
\bibfield{author}{\bibinfo{person}{Krygsman Stephan}, \bibinfo{person}{Dijst
  Martin}, {and} \bibinfo{person}{Arentze Theo}.}
  \bibinfo{year}{2004}\natexlab{}.
\newblock \showarticletitle{Multimodal public transport: an analysis of travel
  time elements and the interconnectivity ratio}. In \bibinfo{booktitle}{{\em
  Transport Policy}}. \bibinfo{publisher}{Elsevier}.
\newblock


\bibitem[\protect\citeauthoryear{Wang, Pedreschi, Song, Giannotti, and
  Barbasi}{Wang et~al\mbox{.}}{2011}]%
        {Wang11}
\bibfield{author}{\bibinfo{person}{D. Wang}, \bibinfo{person}{D. Pedreschi},
  \bibinfo{person}{C. Song}, \bibinfo{person}{F. Giannotti}, {and}
  \bibinfo{person}{A.~L. Barbasi}.} \bibinfo{year}{2011}\natexlab{}.
\newblock \showarticletitle{{Human Mobility, Social Ties and Link Prediction}}.
  In \bibinfo{booktitle}{{\em ACM SIGKDD}}.
\newblock


\bibitem[\protect\citeauthoryear{Wang, Yuan, Lian, Xu, Xie, Chen, and Rui}{Wang
  et~al\mbox{.}}{2015}]%
        {Wang15}
\bibfield{author}{\bibinfo{person}{Y. Wang}, \bibinfo{person}{N.~J. Yuan},
  \bibinfo{person}{D. Lian}, \bibinfo{person}{L. Xu}, \bibinfo{person}{X. Xie},
  \bibinfo{person}{E. Chen}, {and} \bibinfo{person}{Y. Rui}.}
  \bibinfo{year}{2015}\natexlab{}.
\newblock \showarticletitle{{Regularity and Conformity: Location Prediction
  Using Heterogeneous Mobility Data}}. In \bibinfo{booktitle}{{\em ACM
  SIGKDD}}.
\newblock


\bibitem[\protect\citeauthoryear{Xue, Qi, Xie, Zhang, Huang, and Li}{Xue
  et~al\mbox{.}}{2015}]%
        {Xue15}
\bibfield{author}{\bibinfo{person}{A.~Y. Xue}, \bibinfo{person}{J. Qi},
  \bibinfo{person}{X. Xie}, \bibinfo{person}{R. Zhang}, \bibinfo{person}{J.
  Huang}, {and} \bibinfo{person}{Y. Li}.} \bibinfo{year}{2015}\natexlab{}.
\newblock \showarticletitle{{Solving the data sparsity problem in destination
  prediction}}.
\newblock \bibinfo{journal}{{\em Very Large DataBase (VLDB)\/}}
  (\bibinfo{year}{2015}).
\newblock


\bibitem[\protect\citeauthoryear{Xue, Zhang, Zheng, Xie, Huang, and Xu}{Xue
  et~al\mbox{.}}{2013}]%
        {Xue13}
\bibfield{author}{\bibinfo{person}{A.~Y. Xue}, \bibinfo{person}{R. Zhang},
  \bibinfo{person}{Y. Zheng}, \bibinfo{person}{X. Xie}, \bibinfo{person}{J.
  Huang}, {and} \bibinfo{person}{Z. Xu}.} \bibinfo{year}{2013}\natexlab{}.
\newblock \showarticletitle{{Destination prediction by sub-trajectory synthesis
  and privacy protection against such prediction}}. In \bibinfo{booktitle}{{\em
  ACM International Conference on Data Engineering (ICDE)}}.
\newblock


\bibitem[\protect\citeauthoryear{Yuan, Wang, Zhang, Xie, and Sun}{Yuan
  et~al\mbox{.}}{2013}]%
        {Yuan13}
\bibfield{author}{\bibinfo{person}{N.~J. Yuan}, \bibinfo{person}{Y. Wang},
  \bibinfo{person}{F. Zhang}, \bibinfo{person}{X. Xie}, {and}
  \bibinfo{person}{G. Sun}.} \bibinfo{year}{2013}\natexlab{}.
\newblock \showarticletitle{{Reconstructing Individual Mobility from Smart Card
  Transactions}}. In \bibinfo{booktitle}{{\em IEEE International Conference on
  Data Mining (ICDM)}}.
\newblock


\bibitem[\protect\citeauthoryear{Zang and Bolot}{Zang and Bolot}{2011}]%
        {Zang2011}
\bibfield{author}{\bibinfo{person}{Hui Zang} {and} \bibinfo{person}{Jean
  Bolot}.} \bibinfo{year}{2011}\natexlab{}.
\newblock \showarticletitle{Anonymization of Location Data Does Not Work: A
  Large-scale Measurement Study}. In \bibinfo{booktitle}{{\em Proceedings of
  the 17th Annual International Conference on Mobile Computing and Networking}}
  {\em (\bibinfo{series}{MobiCom '11})}. \bibinfo{publisher}{ACM}.
\newblock


\bibitem[\protect\citeauthoryear{Zhang, Zheng, and Yu}{Zhang
  et~al\mbox{.}}{2018a}]%
        {Zhang18_2}
\bibfield{author}{\bibinfo{person}{H. Zhang}, \bibinfo{person}{Y. Zheng}, {and}
  \bibinfo{person}{Y. Yu}.} \bibinfo{year}{2018}\natexlab{a}.
\newblock \showarticletitle{Detecting Urban Anomalies Using Multiple
  Spatio-Temporal Data Sources}.
\newblock \bibinfo{journal}{{\em Journal of the ACM on Interactive, Mobile,
  Wearable and Ubiquitous Technologies\/}} (\bibinfo{year}{2018}).
\newblock


\bibitem[\protect\citeauthoryear{Zhang, Zheng, Qi, Li, Yi, and Li}{Zhang
  et~al\mbox{.}}{2018b}]%
        {Zhang18}
\bibfield{author}{\bibinfo{person}{Junbo Zhang}, \bibinfo{person}{Yu Zheng},
  \bibinfo{person}{Dekang Qi}, \bibinfo{person}{Ruiyuan Li},
  \bibinfo{person}{Xiuwen Yi}, {and} \bibinfo{person}{Tianrui Li}.}
  \bibinfo{year}{2018}\natexlab{b}.
\newblock \showarticletitle{Predicting citywide crowd flows using deep
  spatio-temporal residual networks}.
\newblock \bibinfo{journal}{{\em Artificial Intelligence\/}}
  (\bibinfo{year}{2018}).
\newblock


\bibitem[\protect\citeauthoryear{Zhang, Rossi, and Pavone}{Zhang
  et~al\mbox{.}}{2016}]%
        {Zhang16}
\bibfield{author}{\bibinfo{person}{R. Zhang}, \bibinfo{person}{F. Rossi}, {and}
  \bibinfo{person}{M. Pavone}.} \bibinfo{year}{2016}\natexlab{}.
\newblock \showarticletitle{{Model Predictive Control of Autonomous Mobility
  On-demand Systems}}. In \bibinfo{booktitle}{{\em Robotics and Automation}}.
\newblock


\bibitem[\protect\citeauthoryear{Zhou, Ma, Zhang, Suia, Pei, and
  Moscibroda}{Zhou et~al\mbox{.}}{2016}]%
        {Zhou16}
\bibfield{author}{\bibinfo{person}{M. Zhou}, \bibinfo{person}{M. Ma},
  \bibinfo{person}{Y. Zhang}, \bibinfo{person}{K. Suia}, \bibinfo{person}{S.
  Pei}, {and} \bibinfo{person}{T. Moscibroda}.}
  \bibinfo{year}{2016}\natexlab{}.
\newblock \showarticletitle{{EDUM: Classroom Education Measurements via
  Large-scale WiFi Networks}}. In \bibinfo{booktitle}{{\em ACM International
  Joint Conference on Pervasive and Ubiquitous Computing (Ubicomp)}}.
\newblock


\end{thebibliography}

\end{document}